\documentclass{emulateapj}

\usepackage{color}
\usepackage[usenames,dvipsnames,svgnames,table]{xcolor}
\usepackage{subfigure}
\usepackage{graphicx}
\usepackage{threeparttable}
\usepackage{rotating}
\usepackage{lipsum}
\usepackage{arydshln}
\usepackage{amsmath}
\usepackage{hyperref}
\usepackage{booktabs}
\usepackage{xcolor}
\usepackage{soul}
\usepackage{natbib}
\bibpunct{(}{)}{;}{a}{}{,}

\begin{document}

\title{Emission Line Metallicities from the Faint Infrared Grism Survey and VLT/MUSE}

\author{John Pharo\altaffilmark{1},
        Sangeeta Malhotra\altaffilmark{1,2},
        James Rhoads\altaffilmark{1,2},
        Lise Christensen\altaffilmark{3},
        Steven L. Finkelstein\altaffilmark{4},
        Norman Grogin\altaffilmark{5},
        Santosh Harish\altaffilmark{1},
        Tianxing Jiang\altaffilmark{1},
        Keunho Kim\altaffilmark{1},
        Anton Koekemoer\altaffilmark{5},
        Norbert Pirzkal\altaffilmark{5},
        Mark Smith\altaffilmark{1},
        Huan Yang\altaffilmark{6,1},
        Andrea Cimatti\altaffilmark{7,8},
        Ignacio Ferreras\altaffilmark{9},
        Nimish Hathi\altaffilmark{5},
        Pascale Hibon\altaffilmark{10},
        Gerhardt Meurer\altaffilmark{11},
        Goeran Oestlin\altaffilmark{12},
        Anna Pasquali\altaffilmark{13},
        Russell Ryan\altaffilmark{5},
        Amber Straughn\altaffilmark{2}, and
        Rogier Windhorst\altaffilmark{1}}

\altaffiltext{1}{School of Earth \& Space Exploration, Arizona State University, Tempe, AZ. 85287-1404, USA}
\altaffiltext{2}{NASAs Goddard Space Flight Center, Astrophysics Science Division, Code 660, Greenbelt, MD 20771 USA}
\altaffiltext{3}{1 Dark Cosmology Centre, Niels Bohr Institute, University of Copenhagen, Juliane Maries Vej 30, DK-2100 Copenhagen, Denmark}
\altaffiltext{4}{Department of Astronomy, The University of Texas at Austin, Austin, TX 78712, USA}
\altaffiltext{5}{Space Telescope Science Institute, Baltimore, MD 21218, USA}
\altaffiltext{6}{CAS Key Laboratory for Research in Galaxies and Cosmology, Department of Astronomy, University of Science and Technology of China, China}
\altaffiltext{7}{Department of Physics and Astronomy (DIFA), University of Bologna, Via Gobetti 93/2, I-40129, Bologna, Italy}
\altaffiltext{8}{INAF—Osservatorio Astrofisico di Arcetri, Largo E. Fermi 5, I-50125, Firenze, Italy}
\altaffiltext{9}{Mullard Space Science Laboratory, University College London, Holmbury St. Mary, Dorking, Surrey RH5 6NT, UK}
\altaffiltext{10}{ESO, Alonso de Cordova 3107, Santiago, Chile}
\altaffiltext{11}{International Centre for Radio Astronomy Research, The University of Western Australia, Crawley WA 6009, Australia}
\altaffiltext{12}{Stockholm University, Stockholm, SE-10691, Sweeden}
\altaffiltext{13}{Astronomisches Rechen-Institut, Zentrum fuer Astronomie, Universitaet Heidelberg, Moenchhofstrasse 12-14, D-69120 Heidelberg, Germany}

\begin{abstract}
 We derive direct measurement gas-phase metallicities of $7.4 < 12 + \log(O/H) < 8.4$ for 14 low-mass Emission Line Galaxies (ELGs) at $0.3 < z < 0.8$ identified in the Faint Infrared Grism Survey (FIGS). We use deep slitless G102 grism spectroscopy of the Hubble Ultra Deep Field (HUDF), dispersing light from all objects in the field at wavelengths between 0.85 and 1.15 microns. We run an automatic search routine on these spectra to robustly identify 71 emission line sources, using archival data from VLT/MUSE to measure additional lines and confirm redshifts. We identify 14 objects with $0.3 < z < 0.8$ with measurable O[\textsc{iii}]$\lambda$4363 \AA\ emission lines in matching VLT/MUSE spectra. For these galaxies, we derive direct electron-temperature gas-phase metallicities with a range of $7.4 < 12 + \log(O/H) < 8.4$. With matching stellar masses in the range of $10^{7.9} M_{\odot} < M_{\star} < 10^{10.4} M_{\odot}$, we construct a mass-metallicity (MZ) relation and find that the relation is offset to lower metallicities compared to metallicities derived from alternative methods (e.g.,$R_{23}$, O3N2, N2O2) and continuum selected samples. Using star formation rates (SFR) derived from the $H\alpha$ emission line, we calculate our galaxies' position on the Fundamental Metallicity Relation (FMR), where we also find an offset toward lower metallicities. This demonstrates that this emission-line-selected sample probes objects of low stellar masses but even lower metallicities than many comparable surveys. We detect a trend suggesting galaxies with higher Specific Star Formation (SSFR) are more likely to have lower metallicity. This could be due to cold accretion of metal-poor gas that drives star formation, or could be because outflows of metal-rich stellar winds and SNe ejecta are more common in galaxies with higher SSFR.

\end{abstract}


\section{Introduction}

The identification and study of nebular emission lines in galaxies can provide insight into star formation rates, ionization parameters, and gas-phase metallicities, among other physical parameters. The gas-phase metallicity can be related to star formation and mass growth in galaxies via the mass-metallicity (MZ) relation, an observed correlation between a galaxy's stellar mass and its gas-phase metallicity, and by the Fundamental Metallicity Relation \citep{man10, lar10}, an empirical plane relating the metallicity and the stellar mass to the star formation rate.

These relations have been well-established for local star-forming galaxies \citep{tre04}, which show an increase in gas-phase metallicity as stellar mass increases from $10^{8.5} M_{\odot}$ to $10^{10.5} M_{\odot}$, after which the metallicity flattens. Further surveys have pushed the study of the relation out to higher redshifts, typically finding lower levels of metallicity out to $z \sim 3$ \citep{lil03, mai05, erb06, man09}. For these studies, the gas-phase metallicity is often measured through empirical and theoretical strong line ratio calibrations, such as $R_{23}$ \citep{kk04}, N2O2 \citep{kd02}, and O3N2 \citep{pp04}, using [O\textsc{ iii}], [O\textsc{ ii}], and Balmer-series hydrogen lines (see Table 1 for description of ratios), or via modeling UV indicators including C\textsc{ iii}1907 \citep{amo17}. However, offsets between local and high-redshift galaxies on diagnostic plots such as the Baldwin-Phillips-Terlevich (BPT) diagram \citep{bal81, ste14, san15}, which compares the [O\textsc{ iii}]$\lambda$5007/H$\beta$ line ratio to the [N\textsc{ ii}]$\lambda$6568/H$\alpha$ line ratio, indicate that conditions in the interstellar medium may differ at different redshifts \citep{kew13}. If so, there may be undetected biases in the line ratio calibrations. Some studies have also indicated, however, that the presence of very strong emission lines is itself an indicator of low gas-phase metallicity, regardless of the redshift \citep{fin11, xia12, yan17}. Given these uncertainties and outliers, it is necessary to seek out samples of ELGs for which we can precisely determine the metallicity, and thus better understand its relationship to the other emission properties of galaxies.


\begin{table}
\begin{center}
\caption{Common Strong Line Ratios}
\begin{tabular}{cc}
\tableline
\tableline
Name & Ratio \\
\tableline
N2 & log([N\textsc{ ii}]$\lambda$6584/H$\alpha$) \\
O2 & log([O\textsc{ ii}]$\lambda$3727+3729/H$\beta$) \\
O3 & log([O\textsc{ iii}]$\lambda$4959+5007/H$\beta$) \\
$R_{23}$ & log(([O\textsc{ iii}]$\lambda$4959+5007 + [O\textsc{ ii}]$\lambda$3727+3729)/H$\beta$) \\
N2O2 & N2 - O2 \\
O3N2 & O3 - N2 \\
\tableline
\end{tabular}
\end{center}
\end{table}

A more direct method for measuring the gas-phase metallicity involves the ratio of the auroral [O\textsc{ iii}]$\lambda$4363 \AA\ emission line to the [O\textsc{ iii}]$\lambda$4959,5007 \AA\ lines, which is sensitive to the electron temperature of the ionized gas \citep{izo06, all84, de86, kd02}. A direct measurement of the electron temperature allows for the derivation of abundances with a minimum of other assumptions compared to the more common strong-emission-line diagnostics described above. For example, the $R_{23}$ relation is double-branched, with each $R_{23}$ value corresponding to both a high-metallicity and a low-metallicity solution, requiring additional data or assumptions to break the degeneracy. Consequently, direct-temperature-derived metallicities are more reliable \citep{izo06}. This method is not always practical, as the auroral line is typically quite weak (50-100 times weaker than typical strong lines, per \citet{san17}) and may require stacking spectra to get a reliable signal \citep{am13}, but it provides more accurate metallicity measurements.

In this paper, we describe our systematic search for Emission Line Galaxies (ELGs) in 1D spectra from the Faint Infrared Grism Survey (FIGS). In \S2, we describe the survey and procedures for data collection and reduction. In \S3, we describe the line search method and the flux measurements for confirmed ELGs. In \S4, we detail the measurement of the gas-phase metallicity, and in \S5 we explore the mass-metallicity relation and other properties available from our line measurements. Finally, we summarize in \S6. For this paper we will use $H_0$ = 67.3 km s$^{-1}$ Mpc$^{-1}$ and $\Omega_M = 0.315$, $\Omega_{\Lambda} = 0.685$ \citep{pla15}. All magnitudes are given in the AB system \citep{og83}.

\section{Survey Description and Data}

\subsection{FIGS Observations and Spectral Extraction}

\subsubsection{Survey Description}

\begin{figure}
  \includegraphics[width=0.5\textwidth]{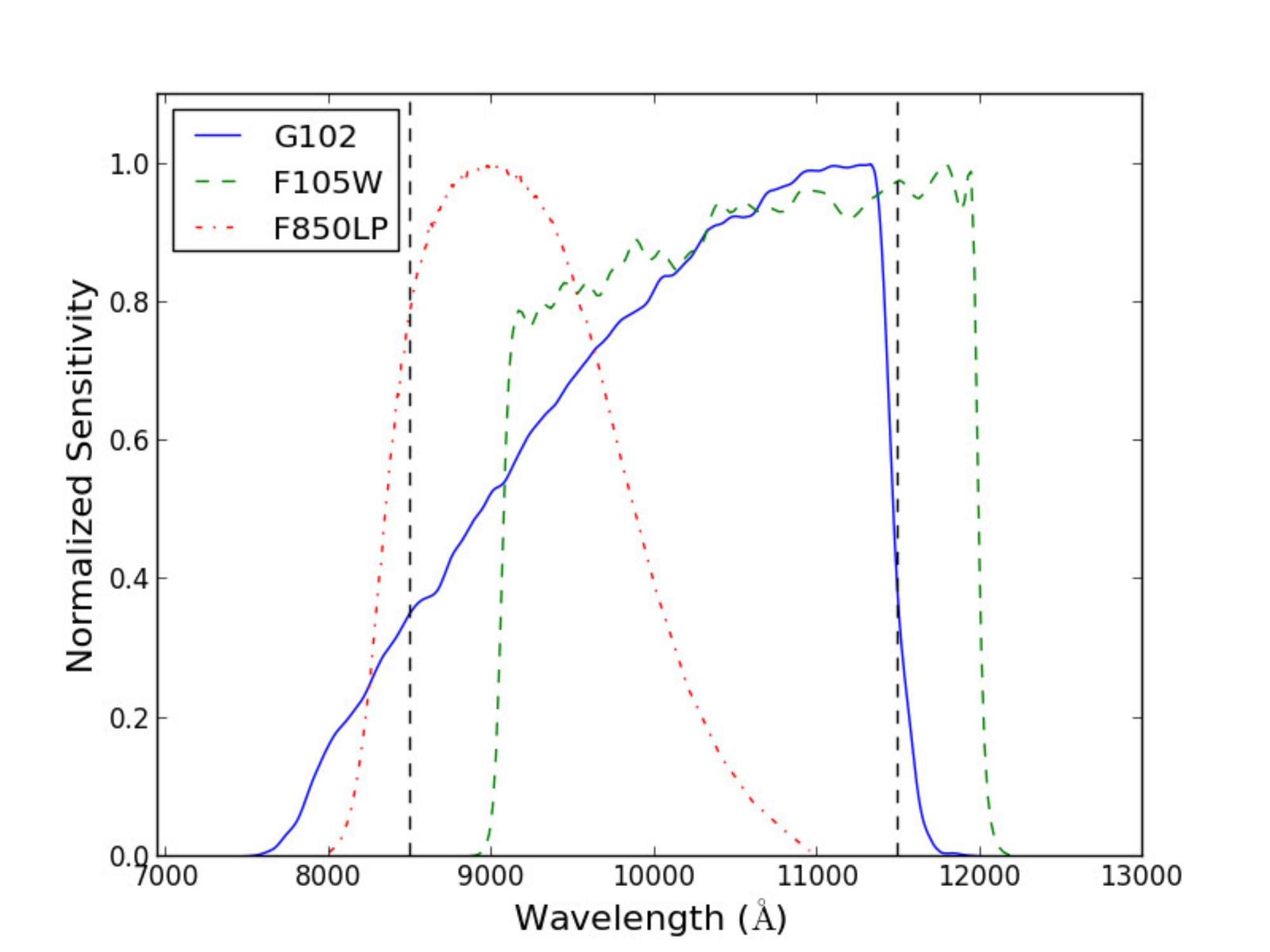}
  \caption{The sensitivity curves for the WFC3/G102 grism, as seen in \citet{kun11}, and the WFC3-F105W and ACS-F850LP filters. The dashed vertical lines show the cutoffs for grism data used in the emission line search. The curves have been normalized to their maximum sensitivity, so this plot gives the sensitivity at each wavelength in terms of its percentage of the peak sensitivity.}
\end{figure}

The Faint Infrared Grism Survey (FIGS, HST/Cycle 22, ID:13779, PI S. Malhotra) used the HST WFC3-G102 (see Figure 1) infrared grism to obtain deep slitless spectroscopy of $\sim$ 6000 galaxies. FIGS achieved 40-orbit depth in 4 fields, designated GN1, GN2, GS1 (UDF), and GS2 (HDF-PAR2) (see Table 2 for coordinates of each field). Objects in each field were observed in 5 different 8-orbit position angles (PAs) in order to mitigate contamination of the spectra by overlapping spectra from nearby objects. Each PA covers a 2.05'x2.27' field of view. The area of coverage in each field from which we derive the usable spectra is given in Table 2, for a total area of 17.7 square arcminutes. 

\begin{table}
\begin{center}
\caption{A description of the four FIGS fields.}
\begin{tabular}{cccc}
\tableline
Field & RA & Dec & Area\footnotemark[1]  \\
\tableline
GN1 & 12:36:41.467 & +62:17:26.27 & 4.51 \\
GN2 & 12:37:31.023 & +62:18:26.91 & 5.06\\
GS1\footnotemark[2] & 03:32:40.951 & --27:46:47.92 & 4.09 \\
GS2\footnotemark[3] & 03:33:06.468 & --27:51:21.56 & 4.02 \\
\tableline
\end{tabular}
\scriptsize {
  \begin{tablenotes}
    \item[]$\rm ^a$ Measured in arcmin$^2$. \\
  \item[]$\rm ^b$ The HUDF. \\
  \item[]$\rm^c$ The HDF Parallel Field. \\
  \end{tablenotes}}
\end{center}
\end{table}

\subsubsection{Spectral Extraction}

In this paper, we used 1D spectra which were generated using the methods described in \citet{pir17}. Here we briefly summarize this process. We reduced FIGS data  in a manner that loosely follows the method used for GRAPES and PEARS, previous HST grism surveys \citep{pir04, xu07, rho09, str09, xia12, pir13}. First, we generated a master catalog of sources from deep CANDELS survey mosaics in the F850LP filter in ACS and the F125W and F160W filters in WFC3 (approximately the z, J, and H bands) \citep{gro11, koe11}. We astrometrically corrected the data to match the absolute astrometry of the GOODS catalogs. The background levels of the grism observations were estimated using a two-components model which include a constant Zodiacal light background as well as a varying HeI light background. Individual spectra were generated using a Simulation Based Extraction (SBE) approach that accounts for spectral contamination from overlapping spectra, as well as allow the use of an optimal extraction approach \citep{hor86} when generating 1D spectra from 2D spectra. We refer the reader to \citet{pir17} for a complete description of these processes. When the extractions were complete, we had an average of $\sim1700$ spectra per field, with a typical $3\sigma$ detection limit of $m_{F105W} = 26$ mag and an emission line sensitivity of $10^{-17}$ ergs cm$^{-2}$ s$^{-1}$.

\subsection{Optical Data}

We supplemented our infrared FIGS spectra with archival high-resolution optical IFU spectra taken with the Multi-Unit Spectroscopic Explorer (MUSE) instrument \citep{bac10} from the Very Large Telescope (VLT). This expands the available spectroscopic wavelength coverage considerably, enabling confirmation of detected emission lines in FIGS via the identification of complementary emission lines at optical wavelengths. These lines also make possible the mass-metallicity results shown in \S 5. We used the publicly available IFU spectra from the MUSE Hubble Ultra Deep Survey \citep{bac17}, a mosaic of nine $1 \times 1$ arcmin$^2$ MUSE fields in the HUDF. In order to extract spectra for emission-line objects in our sample, we applied the following procedure: Using the known sky coordinates for each object, 1-D spectrum was generated by summing up flux within a 2\arcsec\ aperture (centered on the object) at each wavelength slice, across the entire MUSE wavelength range. We extracted FIGS candidate spectra from the reduced MUSE datacube. The MUSE data wavelength coverage extends from 4752 \AA\ to 9347 \AA\ with a spectral resolution of 2.3 \AA, though the sensitivity drops off precipitously at wavelengths lower than 5000 \AA\ and higher than 9200 \AA, so we restrict our usage to between these wavelengths. MUSE has a $3\sigma$ line sensitivity of $\sim 3 \cdot 10^{-19}$ ergs cm$^{-2}$ s$^{-1}$, and thus is deep enough to detect the weak [O\textsc{ iii}]4363 line for FIGS-selected ELGs.

\section{Line Identification and Flux Measurement Methods}

\subsection{Line Identification}

Because we obtained our infrared spectra via slitless grism spectroscopy, there is no pre-selection of ELG candidates via the placement of slits or by broadband magnitude cutoffs. This has the advantage of enabling the detection of ELGs with potentially very low continuum levels, and so might allow for the study of smaller and/or fainter galaxies with nebular line emission. However, this does require an efficient method for selecting ELG candidates from the total sample of FIGS objects. In order to search the $\sim 6000$ FIGS spectra for emission lines, we developed a code to automatically search for and identify peaks in a 1D spectrum. 

First, the continuum flux needs to be estimated at each wavelength element. The G102 grism measures the spectrum every 24.5 \AA, and we use the spectrum from 8500 \AA\ to 11500 \AA. The algorithm iterates over each wavelength element in a given spectrum, estimating the continuum flux at that wavelength and subtracting it. This estimation is accomplished via a median-flux filter, where, given a number of wavelength elements for the width of a prospective line, the algorithm measures the flux in a number of elements outside the guessed line width in both the blue and red wavelength directions. The median flux of all of these points is assumed to represent the continuum there, and is subtracted from that point's flux. This serves to estimate the local value of the continuum while avoiding the influence of the line flux itself or of other features or changes in the spectrum. See Figure 2 for an example continuum-subtracted spectrum.

Next, we calculate the signal-to-noise ratio ($S/N$) at each wavelength with the residual flux and the flux error (determined by the standard deviation in the fluxes selected for measuring the continuum), once more iterating through the list of wavelength elements. The signal is determined by a sum of the fluxes across 5 points centered on the wavelength of the current iteration, and the noise is the same but added in quadrature. Then the algorithm identified the maximum $S/N$; if the ratio exceeds 5, we fit a Gaussian at the central wavelength element, integrate it to get the flux, and subtract the Gaussian from the residual flux. Then we check the next-highest $S/N$, and if it still exceeds 5, the routine repeats until the peak $S/N$ is below the detection threshold.

We run this routine on the individual PA spectra in each field, and record all instances of $S/N > 5$. If the code finds a peak in at least two PAs with centroids at the same or adjacent wavelength elements (24.5 \AA\ in either direction), it declares a detection. In this paper, we focus on only one of the fields, GS1/HUDF, and specifically on candidates with optical data available for line confirmation. A broader catalog of 1D-selected ELGs from FIGS will be explored in a future paper \citep{pha19}. A search for ELGs in the FIGS 2D spectra can be found in \citet{pir18}.

\begin{figure}
  \includegraphics[width=0.5\textwidth]{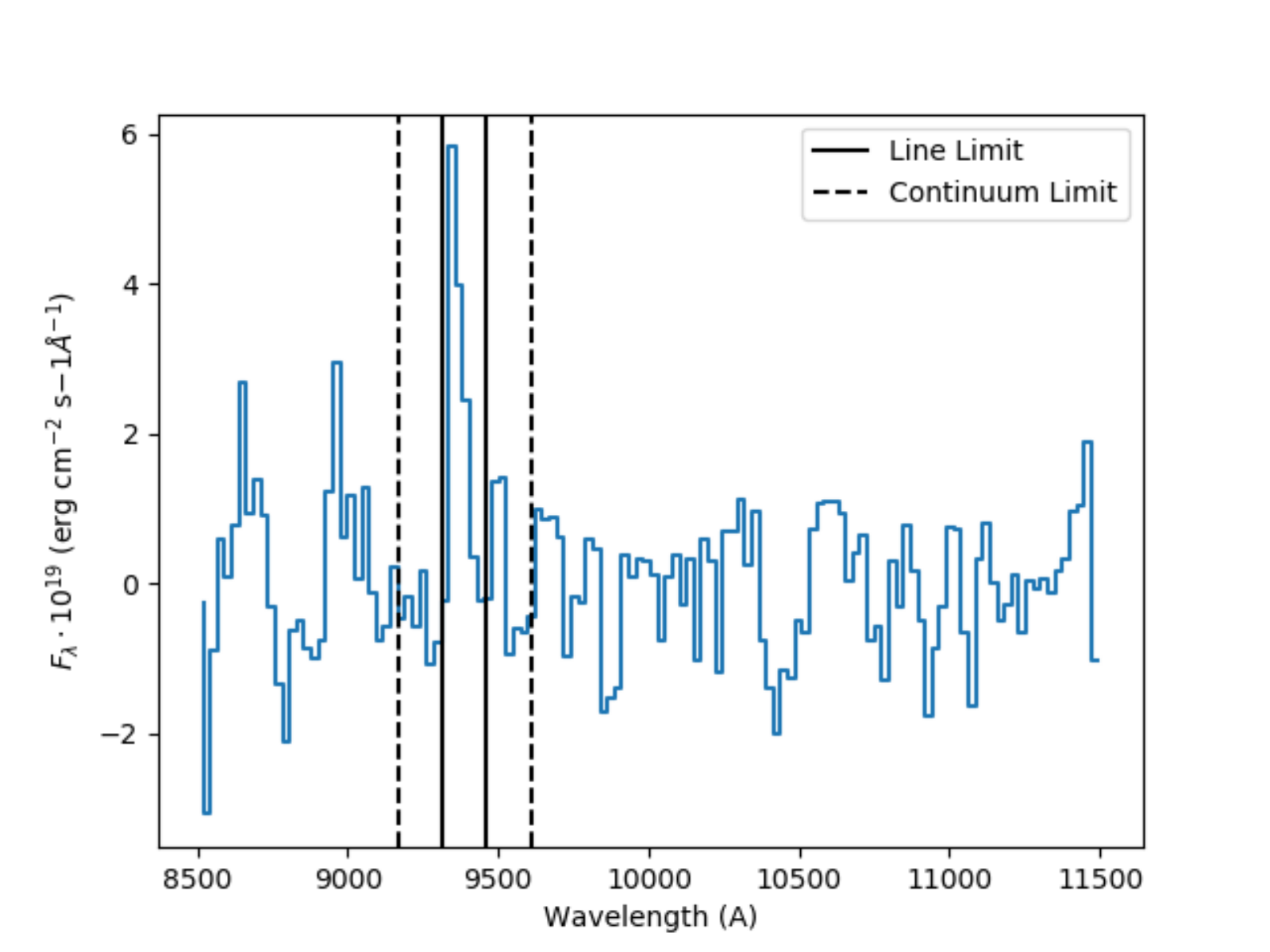}
  \caption{An example of the line-finding routine. This plot shows the continuum-subtracted flux for one PA of FIGS object GS1-2375. This shows an iteration of the line-finding routine when centered at 9388 \AA. The routine sums the flux of the pixels within the solid black lines, which is considered to be the candidate line flux. Then, the continuum flux is estimated from the median flux of the pixels between the solid and dashed lines. This continuum is subtracted from the line flux, and the S/N is calculated.}
\end{figure}

In the GS1/HUDF field, where we have matching optical MUSE spectra, this method produces 137 candidate emission line objects. Of these, 131 had matches in the MUSE source catalog within 1 arcsecond of separation. Using our FIGS redshift catalogs \citep{pha18}, we matched the candidate list with their redshifts and sorted the candidates according to the likely spectral emission line at that redshift. We use the wavelength of the peak S/N pixel to get an approximate rest-frame line centroid. We also compared our candidates with emission lines identified in the GRAPES Survey with the HST ACS G800L grism \citep{xu07}.

When identifying the FIGS-spectra emission lines, we considered common strong lines: Ly$\alpha$, H$\alpha$, H$\beta$, [Mg\textsc{ ii}]$\lambda$2798 \AA, [O\textsc{ iii}]$\lambda$5007 \AA, and [O\textsc{ ii}]$\lambda$3727 \AA, though fainter lines could often be seen in the higher-resolution MUSE spectra. We identified the FIGS lines by determining the ratio of observed wavelengths between two detected emission lines, a fixed property for a given pair of emission lines that is not dependent on the redshift. If no other emission line was detected, we identified the line with the object's photometric redshift. This produced 32 [O\textsc{ iii}]$\lambda$5007 \AA\ candidates ($z \simeq 0.82 - 1.35$), 22 H$\alpha$ candidates ($z \simeq 0.30 - 0.80$), and 17 [O\textsc{ ii}]$\lambda$3727 \AA\ candidates ($z \simeq 1.30 - 2.0$). The remaining detections were ruled out as contamination or some other false detection (e.g., due to a sharp change in continuum slope) upon visual inspection, were ruled out by other line detections in MUSE, or had bad redshift calculations. These tended to be among the faintest objects, which are more susceptible to contamination from nearby objects. In order to cast a wide net for ELGs, we did not impose a continuum magnitude limit on the search, relying on visual inspection and other spectra to confirm our detections. 

Of the 32 [O\textsc{ iii}] candidates from FIGS, 11 were confirmed by inspecting matching MUSE spectra, which means we either measured the same line in the region of overlapping wavelength coverage (8500 - 9300 \AA), or measured a second line which produced a wavelength ratio consistent with an [O\textsc{ iii}]$\lambda\lambda$4959+5007-[O\textsc{ ii}]$\lambda\lambda$3727+3729 pair. However, due to the presence of atmospheric emission lines in MUSE and the fact that we cannot know [O\textsc{ ii}] strength just from [O\textsc{ iii}] detection, lack of a clear [O\textsc{ ii}] detection does not rule out the line being [O\textsc{ iii}]. Matches with the GRAPES line list confirmed an additional 7 candidates, leaving 13 unconfirmed (though the line ID is still implied by the redshift) and 1 confirmed to be H$\alpha$.

We used a similar process for the FIGS H$\alpha$ candidates, of which 15 were confirmed by MUSE, 3 by GRAPES, and 5 were unconfirmed except by photometric redshift (photo-z). For [O\textsc{ ii}], MUSE can only reliably provide confirmation if the [O\textsc{ ii}] line is in the overlap region, or if another feature (eg, 4000 \AA\ break), happens to be visible. Only 4 could be confirmed this way, and 1 more from GRAPES, leaving 12 candidates unconfirmed except by photo-z. See Table 3 for a summary of these results.

\begin{table}
\begin{center}
\caption{The GS1/HUDF emission line candidates by identification}
\begin{tabular}{ccccccc}
\tableline
ID & Initial & MUSE & GRAPES & Photo-z & Wrong & Total \\
\tableline
[O\sc{ iii}] & 32 & 11 & 7 & 13 & 1\footnotemark[1] & 31 \\
H$\alpha$ & 22 & 15 & 3 & 5 & 0 & 23 \\
{[O\sc{ ii}]} & 17 & 4 & 1 & 12 & 0 & 17 \\
\end{tabular}
\scriptsize {
  \begin{tablenotes}
  \item[]$\rm ^a$ Later confirmed to be $H\alpha$. \\
  \end{tablenotes}}
\end{center}
\end{table}

\subsection{Flux Measurement}

We calculated the emission line fluxes for all of the emission line candidates, regardless of their confirmation status. Beginning with the brightest FIGS line in the spectrum (H$\alpha$, [O\textsc{ iii}], or [O\textsc{ ii}], depending on the candidate line ID), we performed a Gaussian fit using the Kapteyn Package (Terlouw and Vogelaar, 2015) at the wavelength of the peak in each PA where there was a 5$\sigma$ detection, allowing the Gaussian amplitude and sigma to be free parameters with an initial guess based on the peak flux. The centroid was allowed to vary between the adjacent pixels in order to determine the best-fit line center. We interpolated a Gaussian function from the fit, from which we derived the total line flux. Once all PAs for a single object and line had been fit, we averaged the individual fluxes and propagated the individual errors to get the final line measurement.

Once the primary line fit was completed, we recalculated the redshift based on the line center and used this new redshift measurement to predict the locations of other lines. The Gaussian function representing the previous fit was subtracted from the flux, and then we attempted to fit the next line. We repeated this process for any common, strong emission lines within the wavelength coverage. In the FIGS spectra, the only non-primary line detected with any significance was H$\beta$. If an object had matching MUSE spectra, we applied the same process there as well. Wes estimated the total flux errors based on the propagation of errors in the Gaussian fit parameters, which the Kapteyn fitting package determined in part based on flux errors in the constituent pixels.

For the 18 objects where the H$\alpha$ emission line was detected in FIGS and for which a matching optical spectrum was available, we measured the extinction via the Balmer decrement. All the galaxies for which we later derive $T_e$-based metallicities are included in this set. In order to correct for stellar absorption of the Balmer lines, we follow the procedure used in \citet{ly14}, which covers objects a similar redshift. For objects without measurements in both $H\alpha$ and $H\beta$, we applied an extinction correction using the \citet{cal00} reddening model, following the procedure used in a study of ELGs with $R_{23}$ at comparable redshift \citep{xia11}. The full catalog of flux measurements is listed in Table 6.

\subsection{Line Comparisons in FIGS and MUSE}

\begin{figure}
  \includegraphics[width=0.5\textwidth]{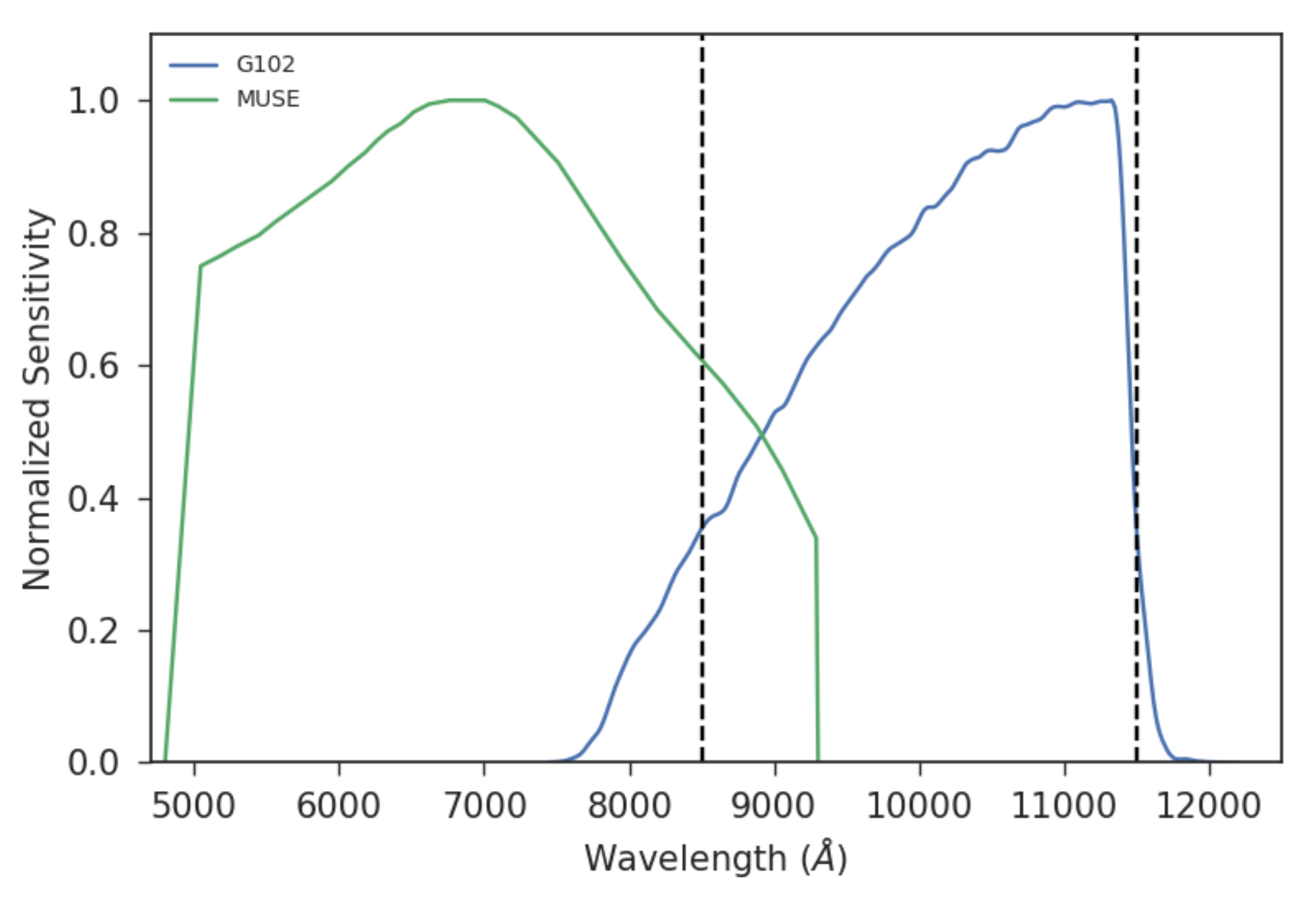}
  \caption{The sensitivity curve for the WFC3/G102 grism \citep{kun11}, and for MUSE \citep{bac10}. The dashed vertical lines show the cutoffs for grism data used in the emission line search. The curves have been normalized to their maximum sensitivity.}
\end{figure}

\begin{figure}
  \includegraphics[width=0.5\textwidth]{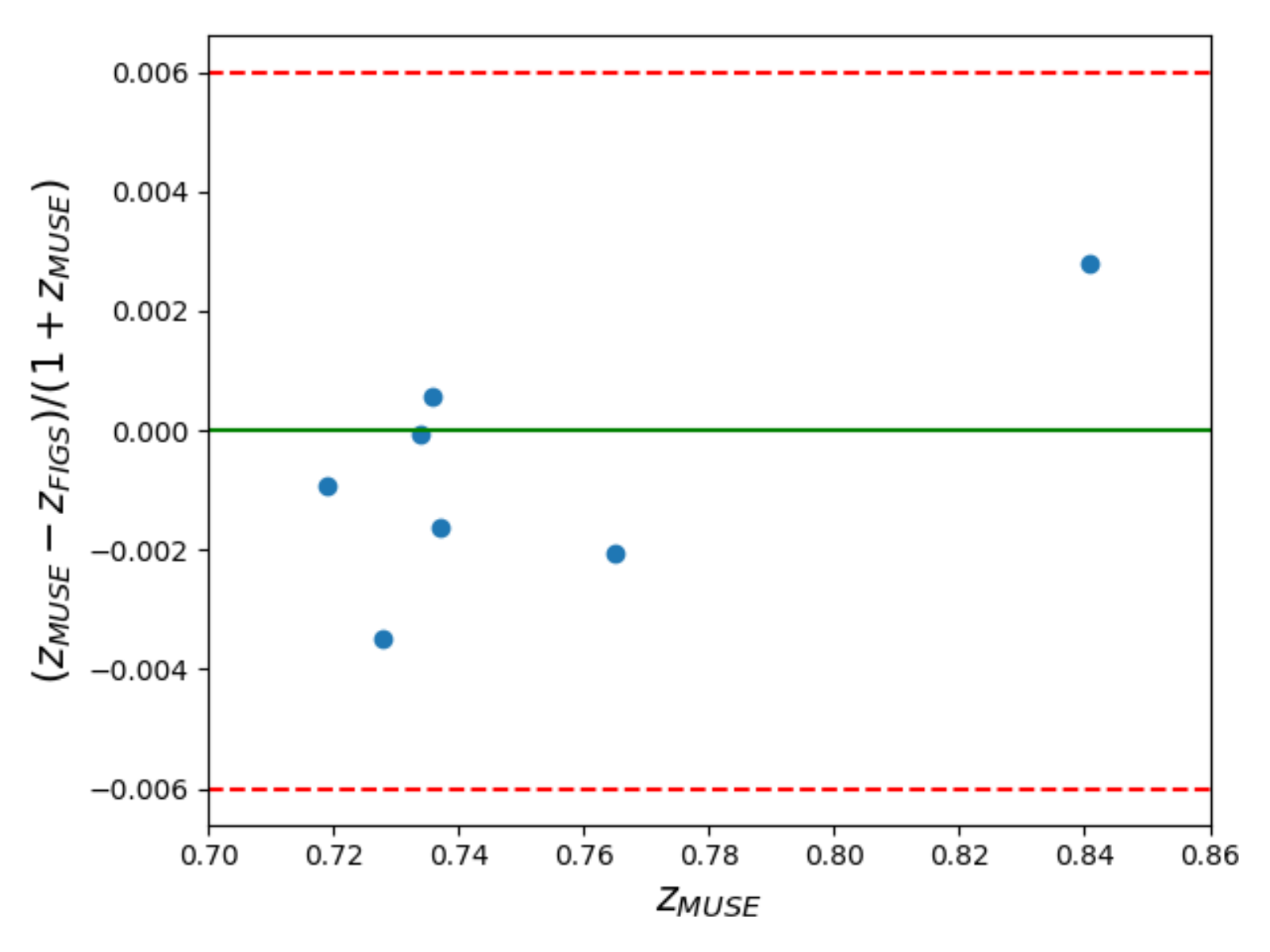}
  \caption{The differences in redshift calculated from the FIGS emission lines and the MUSE emission lines are shown here in blue, with the green line showing a difference of 0. The dashed red lines give the bounds of the RMS wavelength error from (Xia el at. 2011) between the ACS grism and the LDSS3 spectrograph. We measure and RMS redshift difference of $\sigma_z = 0.002$.}
\end{figure}

In addition to using emission lines in optical spectra to confirm line detections in FIGS, the measurement of additional line fluxes for an ELG makes it possible to measure gas-phase metallicities, but it is first necessary to check the consistency of the flux measurements between the two sources of spectra. We were able to check this by looking at emission lines that appeared in both the G102 and MUSE spectra. For emission lines observed between 8500 \AA\ and 9200 \AA\ (See Figure 3), where the throughput of both instruments is good, we were usually able to measure the line flux in both FIGS and MUSE. This provided the opportunity to compare line measurements between the space-based HST/WFC3 and the ground-based VLT/MUSE instruments. In Table 4, we show the flux measurements of the six matching objects, where the [O\textsc{ iii}]$\lambda\lambda$5007,4959 were measured. The matching fluxes are within the measured $1\sigma$ flux errors for four of the seven objects, including the two which are part of the later analysis in this paper. Object 2654 was primarily detected by the H$\alpha$ line, and one PA of the FIGS spectra contains non-removed contamination at the predicted location of the [O\textsc{ iii}] line, which skewed its average flux measurement high. Removing this one PA from the flux measurement brings the FIGS spectra flux into agreement with what we measure in MUSE, bringing the number of well-matched spectra to five of seven objects. 

Because so few objects have an emission line appear in both spectra, it is difficult to judge whether any systematic offset is present from the few where the flux differs. Examining these cases does, however, suggest some possible causes for difference in FIGS and MUSE flux due to contamination or other artifacts, which we checked for visually in our further results. We examined the individual PAs for the 14 objects used in the mass-metallicity analysis to search for any unnoticed contamination that could affect the FIGS lines as with Object 2654, or for any other issues. We discovered no such contamination in any lines required for the metallicity measurement. Object 1299 possibly suffers from oversubtraction of the H$\alpha$ line. However, this object is detected in 5 PAs, so the effect is small.

We also compared the redshifts derived individually from the FIGS line detection and the MUSE line. We calculated the redshift of each object in Table 4 based on the best-fit central wavelength of the line fit for each spectrum. The differences are plotted as a function of MUSE redshift in Figure 4. We find a root-mean-square (RMS) redshift difference ($\Delta z /(1 + z)$) between the two sources of spectra of $\sigma_z = 0.002$. \citet{xia11} found an RMS of $\sigma_z = 0.006$ when comparing redshifts derived from the HST ACS PEARS grism survey and from Magellan's LDSS-3 spectrograph ($R = 100$, as compared to $R = 210$ for G102). This indicates FIGS has a wavelength calibration of at least comparable accuracy to PEARS, given the spectral resolution in each.

\begin{table*}
\begin{center}
\caption{FIGS-MUSE Objects}
\begin{tabular}{ccccccc}
\tableline
FIGS ID & RA & Dec & $\lambda$ (\AA) & z & FIGS Flux\footnotemark[1] & MUSE Flux \footnotemark[1]\\
\tableline
1467 & 53.151047 & -27.777309 & 8735 & 0.736 & 347.7 $\pm$ 27.1 & 365.3 $\pm$ 36.2 \\
1689 & 53.162483 & -27.780346 & 8615 & 0.719 & 890.6 $\pm$ 39.0 & 860.9 $\pm$ 56.7 \\
1851 & 53.152782 & -27.782698 & 8855 & 0.766 & 967.2 $\pm$ 58.4 & 1007.1 $\pm$ 97.3 \\
2560 & 53.184158 & -27.792637 & 8687 & 0.738 & 1821.9 $\pm$ 208.1 & 2041.8 $\pm$ 121.9 \\
2654 & 53.182205 & -27.793993 & 8687 & 0.735 & 315.0\footnotemark[2] $\pm$ 45.5 & 352.1 $\pm$ 50.6 \\
8178 & 53.187664 & -27.783779 & 8663 & 0.734 & 176.6 $\pm$ 28.7 & 94.3 $\pm$ 34.3 \\
\end{tabular}
\scriptsize {
  \begin{tablenotes}
  \item[]$\rm ^a$ Measured in $10^{-19}$ erg/cm$^2$/s. \\
  \item[]$\rm ^b$ Flux calculated after removing one PA for uncorrected contamination, which significantly altered the flux average. \\
  \end{tablenotes}}
\end{center}
\end{table*}

\section{Metallicity Measurements}

\begin{figure*}
\begin{tabular}{cc}
\includegraphics[width=0.5\textwidth]{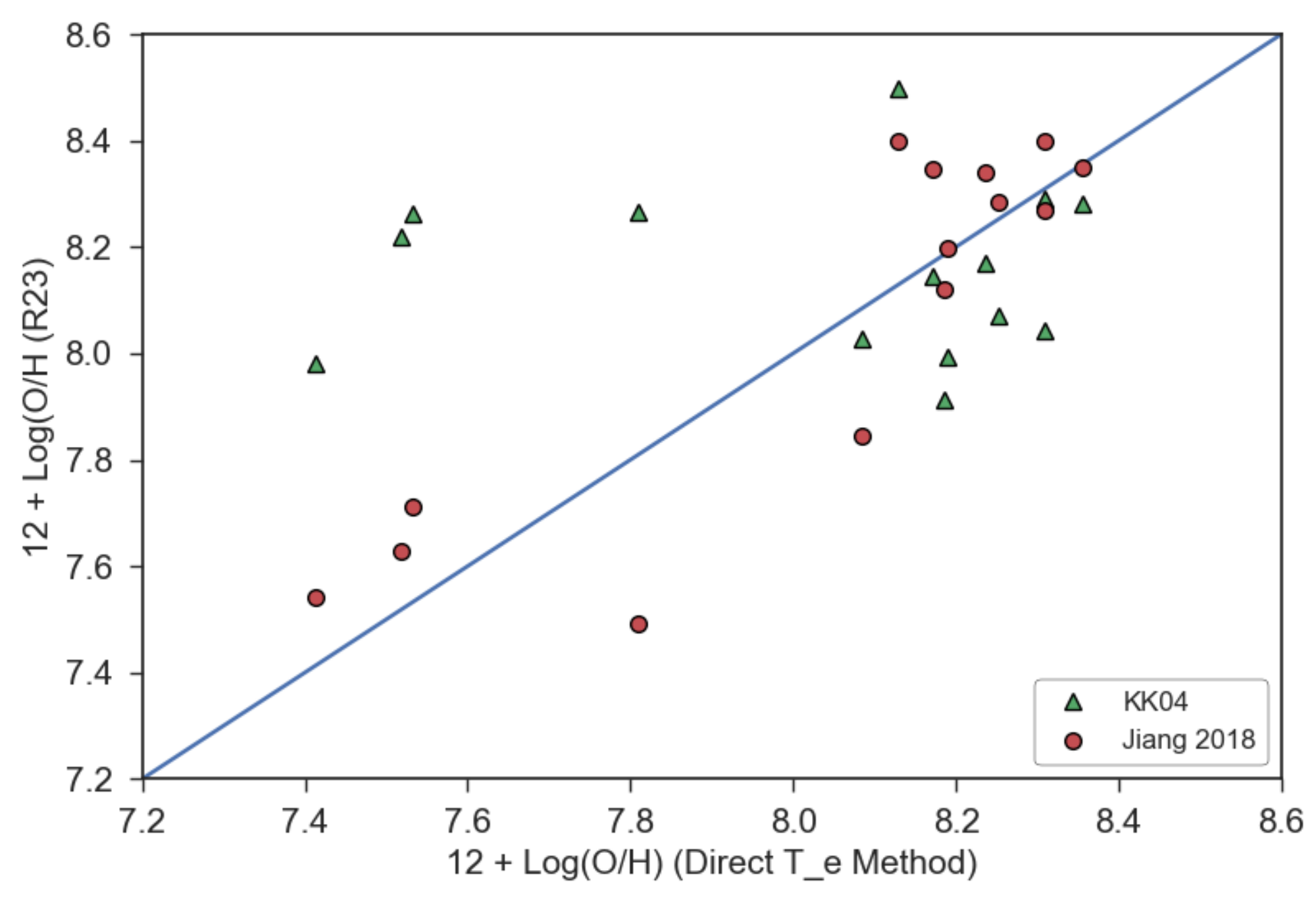} & \includegraphics[width=0.5\textwidth]{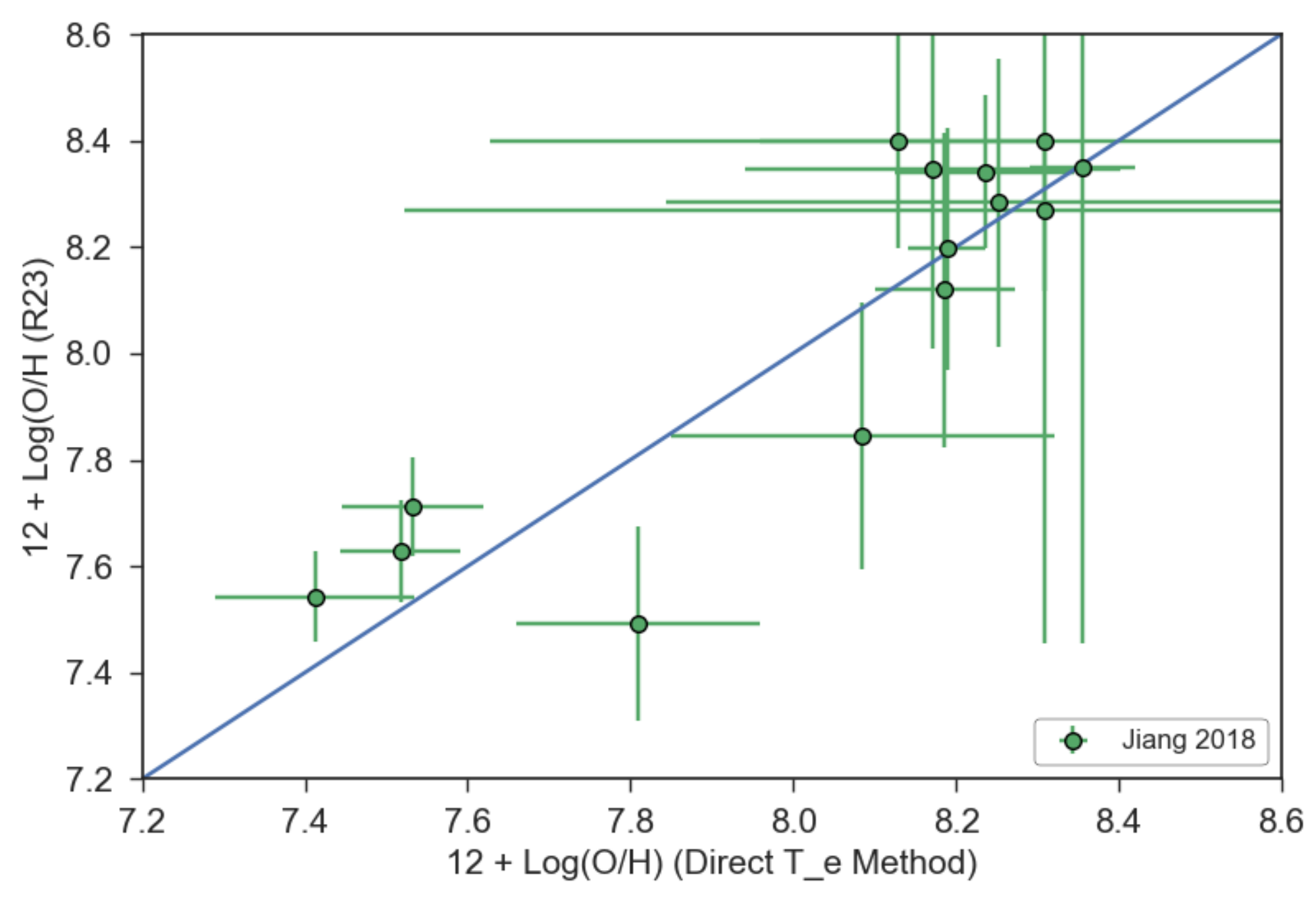}
\end{tabular}
\caption{Left: $R_{23}$ metallicities compared with $T_e$-derived metallicities, using both the KK04 \citep{kk04} parameterization (green triangles) and new calibration based on Green Peas (Jiang et al. 2018) (red circles). Both $R_{23}$ methods place all of the FIGS objects on the lower branch, but the KK04 parameterization tends to overestimate the lowest-metallicity objects, while the Green Pea calibration reduces the scatter considerably. Right: The comparison of the calibrated $R_{23}$ metallicities with $T_e$ metallicities including errors.}
\end{figure*}

Strong nebular emission lines provide the means to measure the gas-phase oxygen abundance in a galaxy, which serves as a proxy for the metallicity. For objects with a [O\textsc{ iii}]$\lambda$4363 \AA\ auroral line detected at $S/N \geq 3$, we used the direct metallicity measurement described in \citet{ly14}, based on the empirical relations in \citet{izo06}. This method first estimates the [O\textsc{ iii}]4363 electron temperature ($T_e$) based on the nebular-to-auroral flux ratio:

\begin{equation}
\log \left( \frac{[O\textsc{ iii}]\lambda \lambda 4959,5007}{[O\textsc{ iii}] \lambda 4363} \right) = \frac{1.432}{t_3} + \log C_T
\end{equation}

where $t_3 = T_e([O\textsc{ iii}])/10^4$ K, and

\begin{equation}
  C_T = (8.44 - 1.09t_3 + 0.5t_3^2 - 0.08t_3^3) \cdot \frac{1 + 0.0004x}{1 + 0.044x}
\end{equation}

where $x = 10^{-4}n_et_3^{-0.5}$ and $n_e$ is the electron density (cm$^{-3}$). Since we are unable to resolve the $[S\textsc{ ii}]\lambda \lambda 6717,6732$ doublet in FIGS, and it is too red to appear in MUSE spectra, we cannot directly measure $n_e$, but $C_T$ is only strongly dependent on $n_e$ in the high-density regime ($n_e > 10^4$ cm$^{-3}$), where $n_e$ is large enough for the $x$ term to be important. We tested the temperature calculation with $n_e = 10, 100, 1000$ cm$^{-3}$ using a range of measured line ratios from \citet{ly14}, and the resulting temperatures were virtually identical for the different density measurements. Thus, we can safely adopt the assumption of \citet{ly14} that $n_e \approx 100$ cm$^{-3}$ for our calculations.

With the temperature estimated, the ionic abundances of oxygen can be determined from the line ratios $[O\textsc{ ii}]\lambda \lambda 3726,3729/H\beta$ and $[O\textsc{ iii}]\lambda \lambda 4959,5007/H\beta$:

\begin{equation}
  12 + \log \left( \frac{O^+}{H^+} \right) = \log \left( \frac{[O\textsc{ ii}]}{H\beta} \right) + 5.961 + \frac{1.676}{t_2}
\end{equation}

\begin{equation}
  12 + \log \left( \frac{O^{++}}{H^+} \right) = \log \left( \frac{[O\textsc{ iii}]}{H\beta} \right) + 6.200 + \frac{1.251}{t_3}
\end{equation}

where $t_2$ is the [O\textsc{ ii}] electron temperature, assuming a two-temperature model $t_2 = T_e ([O\textsc{ ii}])/10^4$ K $= -0.577 + t_3(2.065 - 0.498t_3)$ from \citet{izo06}. In nebular regions, oxygen ions in ionization states other than $O^+$ and $O^{++}$ make up a negligible fraction of the population, so the total oxygen abundance can be determined from

\begin{equation}
  \frac{O}{H} = \left( \frac{O^+}{H^+} \right) + \left( \frac{O^{++}}{H^+} \right)
\end{equation}

14 objects have sufficient [O\textsc{ iii}]4363 signal ($S/N \geq 3$) in the MUSE optical spectra, as well as the other requisite [O\textsc{ iii}], [O\textsc{ ii}], and H$\beta$ lines from FIGS and MUSE, to perform this direct metallicity measurement. We refer the reader to Tables 5 and 6 for the measured line fluxes and spectroscopic sources for individual ELGs. We summarize metallicity and electron temperature measurements for these objects in Table 5. For one of these objects, FIGS ID 2560, we observed a strong peak at the location of the [O \textsc{i}]6300 emission line in one PA, a possible indicator of Seyfert or LINER properties. By consulting line ratio diagnotics in \citet{kew06}, our measurements match the characteristics of a Seyfert galaxy, which could explain the very high temperature measurement, and could skew the metallicity calculation if the [O\textsc{ i}] line is real.

For ELGs without a significant [O\textsc{ iii}]4363 detection, we computed metallicities iteratively using the $R_{23}$ diagnostic \citep{pag79}, given by the ratio $R_{23} = ([O\textsc{ ii}] + [O\textsc{ iii}])/H\beta$. We tested the effectiveness of this method compared to the direct measurement by calculating metallicities using both methods for the 14 objects where this was possible. We found some significant disagreement in the results between the two, particularly at low metallicity, as can be seen in Figure 5. This is not unusual: \citet{ke08} shows that different metallicity diagnostics can produce different measurements of oxygen abundance with a scatter of up to 0.7 dex. However, \citet{chr12} has demonstrated that using an $R_{23}$ calibration with a correction for the ionization parameter based on the [O\textsc{ iii}]/[O\textsc{ ii}] ratio \citep{pt05} agreed well with direct metallicities of $z \sim 2$ galaxies. To address this, we applied a new empirical $R_{23}$ calibration with an [O\textsc{ iii}]/[O\textsc{ ii}] ratio correction, based on a sample of 800 ``green pea" galaxies at $0.011 < z < 0.411$ with reliable direct metallicity measurements (Jiang et al. 2018). This new calibration reduced the scatter between $T_e$-derived metallicities and $R_{23}$-derived metallicities, as can be seen in Figure 5, demonstrating that we could obtain reliable metallicity measurements using $R_{23}$. Thus, we were able to add 8 additional objects to our metallicity sample via the calibrated $R_{23}$ method.

Error measurements for the metallicities are obtained via the propagation of the initial flux errors through the electron temperature calculation combined with error introduced by the extinction correction, and are reported at $1\sigma$. Electron temperature errors are determined by the errors of the line fluxes going into the [O\textsc{ iii}] line ratio: [O\textsc{ iii}]5007,4959 and [O\textsc{ iii}]4363. These are summarized in Table 5.

\begin{table*}
\begin{center}
\caption{FIGS $T_e$ Metallicities}
\begin{tabular}{cccccc}
\tableline
FIGS ID & 12+Log(O/H) & $Log(M_{\star})$ & $\mathbf{Log(T_e/K)}$ & SFR($M_{\odot}$/yr) & z \\
\tableline
950 & $7.81\pm0.15$ & 8.92 & $4.46\pm0.09$ & $0.71\pm0.18$ & 0.678 \\
1016 & $8.25\pm0.23$ & 9.12 & $4.01\pm0.10$ & $0.17\pm0.15$ & 0.622 \\
1103 & $8.17\pm0.23$ & 9.80 & $4.03\pm0.06$ & $0.59\pm 0.23$ & 0.334 \\
1171 & $7.52\pm0.07$ & 8.58 & $4.35\pm0.09$ & $0.76\pm0.13$ & 0.606 \\
1295 & $8.19\pm0.05$ & 9.74 & $4.00\pm0.08$ & $1.41\pm0.17$ & 0.420 \\
1299 & $8.31\pm0.68$ & 9.87 & $4.02\pm0.13$ & $0.59\pm0.76$ & 0.622 \\
1392 & $8.19\pm0.09$ & 9.87 & $4.01\pm0.08$ & $3.18\pm1.27$ & 0.668 \\
1689 & $7.41\pm0.12$ & 8.20 & $4.34\pm0.08$ & $0.65\pm0.18$ & 0.719 \\
2168 & $7.53\pm0.09$ & 7.99 & $4.38\pm0.05$ & $0.12\pm0.02$ & 0.468 \\
2378 & $8.09\pm0.24$ & 10.06 & $4.08\pm0.05$ & $2.92\pm1.20$ & 0.436 \\
2517 & $8.13\pm0.17$ & 9.64 & $4.09\pm0.07$ & $5.99\pm2.13$ & 0.459 \\
2560\footnotemark[1] & $8.24\pm0.11$ & 9.88 & $4.51\pm0.07$ & $5.61\pm1.01$ & 0.738 \\
2783 & $8.31\pm0.79$ & 7.88 & $4.03\pm0.13$ & $0.06\pm0.13$ & 0.532 \\
4198 & $8.36\pm0.06$ & 10.44 & $4.00\pm0.14$ & $3.10\pm0.39$ & 0.669
\end{tabular}
\begin{tablenotes}
For object coordinates and line fluxes, see Table 6. Stellar mass error is $\leq0.1$ dex. See \S5.1 for discussion. \\
\item[]$\rm ^a$ Object 2560 has possible indicators of being a Seyfert galaxy. See \S 4 for details.
\end{tablenotes}
\end{center}
\end{table*}

\section{Results and Discussion}

\subsection{Mass-Metallicity Relation}

We obtained stellar masses from the catalogs in \citet{sant15}, hereafter S15. S15 presented a series of mass catalogs derived from CANDELS photometry (UV to through mid-IR in GOODS-S) and redshifts. The catalogs were computed using a variety of stellar mass codes and a range of preferred modeling parameters. We considered only the mass catalogs whose fits included contributions from nebular emission, which restricted our choice to four of the mass catalogs presented by S15. We use the mass values from one of these, their $6a_{\tau}NEB$ method, which is fit to BC03 templates \citep{bc03} using a Chabrier Initial Mass Function (IMF) and includes the widest range of considered metallicities out of the four methods that consider nebular emission in their SED fits. S15 do not provide individual estimates of the mass error, but did investigate the distributions of mass estimates as compared to the median masses from the list of mass catalogs. They quantified the typical deviation from the median mass with the distribution's semi interquartile range, which they found to be usually less than 0.1 dex, giving a reasonable upper bound on the mass uncertainty.

We matched the S15 catalog with our 22 objects with $T_e$ or $R_{23}$ metallicity measurements within an angular separation of 1 arsecond and confirmed that the CANDELS redshifts provided by the S15 catalog matched the line-derived redshifts for the objects. Then we produced a relation between the stellar mass and the gas-phase oxygen abundance for the 14 objects with $T_e$-derived metallicities, as can be seen in Figure 6. This subsample has a median redshift of $z = 0.614$. We parameterize the FIGS mass-metallicity relation with a quadratic function of the form

\begin{equation}
  12 + \log \left( \frac{O}{H} \right) = A + Bx + Cx^2
\end{equation}

where $x = \log(M_{\star}/M_{\odot}) - 10$. We use a Python function, \textit{curve-fit} from the SciPy package \citep{jon01}, to perform a least squares fit of the FIGS data to this parameterization. The MZ relation is best fitted by

\begin{equation}
  12 + \log \left( \frac{O}{H} \right) = 8.240 + 0.367x - 0.018x^2
\end{equation}

The $1\sigma$ errors in the parameters are determined from the diagonal of the covariance matrix, which gives $\sigma_A = 0.033$, $\sigma_B = 0.089$, and $\sigma_C = 0.051$. We estimate the uncertainty in the fit by performing a Monte Carlo simulation at each stellar mass in the range of the fit (1000 points between $\log(M_{\star}/M_{\odot}) = 7.70$ and 10.44), assuming a Gaussian distribution around these errors. For each mass point, the fit parameters are sampled 10000 times, and the standard deviation of the result is used to estimate the $1\sigma$ uncertainty in the fit. This is represented by the shaded region in Figure 6.

\begin{figure*}
\centerline{\includegraphics{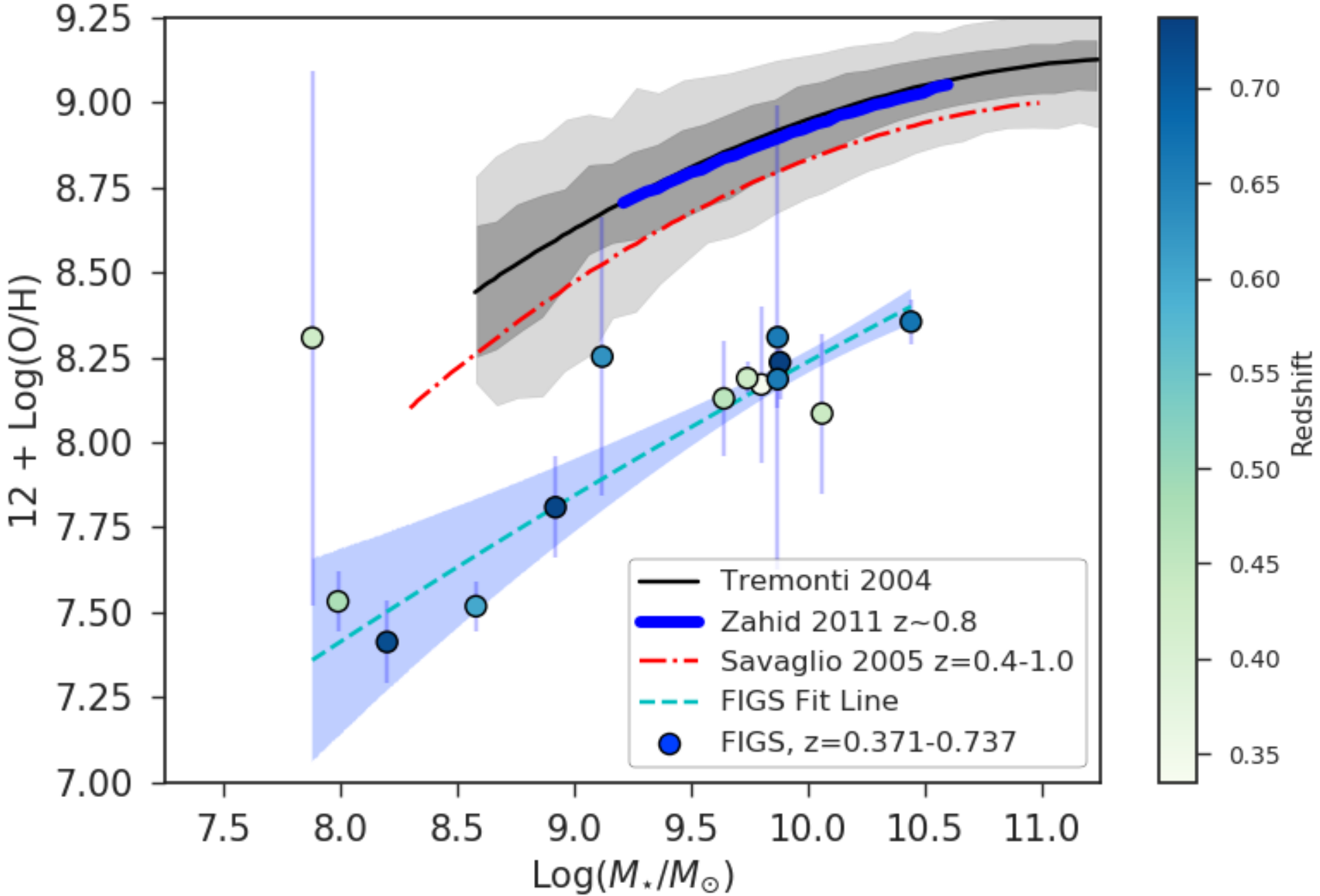}}
\caption{The MZ relation between the stellar masses as measured by \citet{sant15} and the gas-phase oxygen abundances for FIGS objects as measured by $[O\textsc{ iii}]\lambda 4363$ in matching MUSE spectra. The FIGS objects are given by circles with errorbars in metallicity, and are shaded by redshift, with a median of $z=0.614$. The black line and contours represent local SDSS galaxies as measured in \citet{tre04}. The thick blue, solid line represents the $z \sim 0.8$ upper-branch $R_{23}$ metallicities from \citet{zah11}, and the red, dot-dash line represents the $0.4 < z < 1.0$ upper-branch $R_{23}$ metallicities from \citet{sav05}. The blue dashed line is the non-linear least squares fit to the FIGS objects, using SciPy's curve-fit function (See Equation 7 for parameters) \citep{jon01}. The blue shaded region is the $1 \sigma$ uncertainty of the fit as measured from a Monte Carlo of the $1 \sigma$ uncertainties in the fit parameters.}
\end{figure*}

\begin{figure*}
  \centerline{\includegraphics{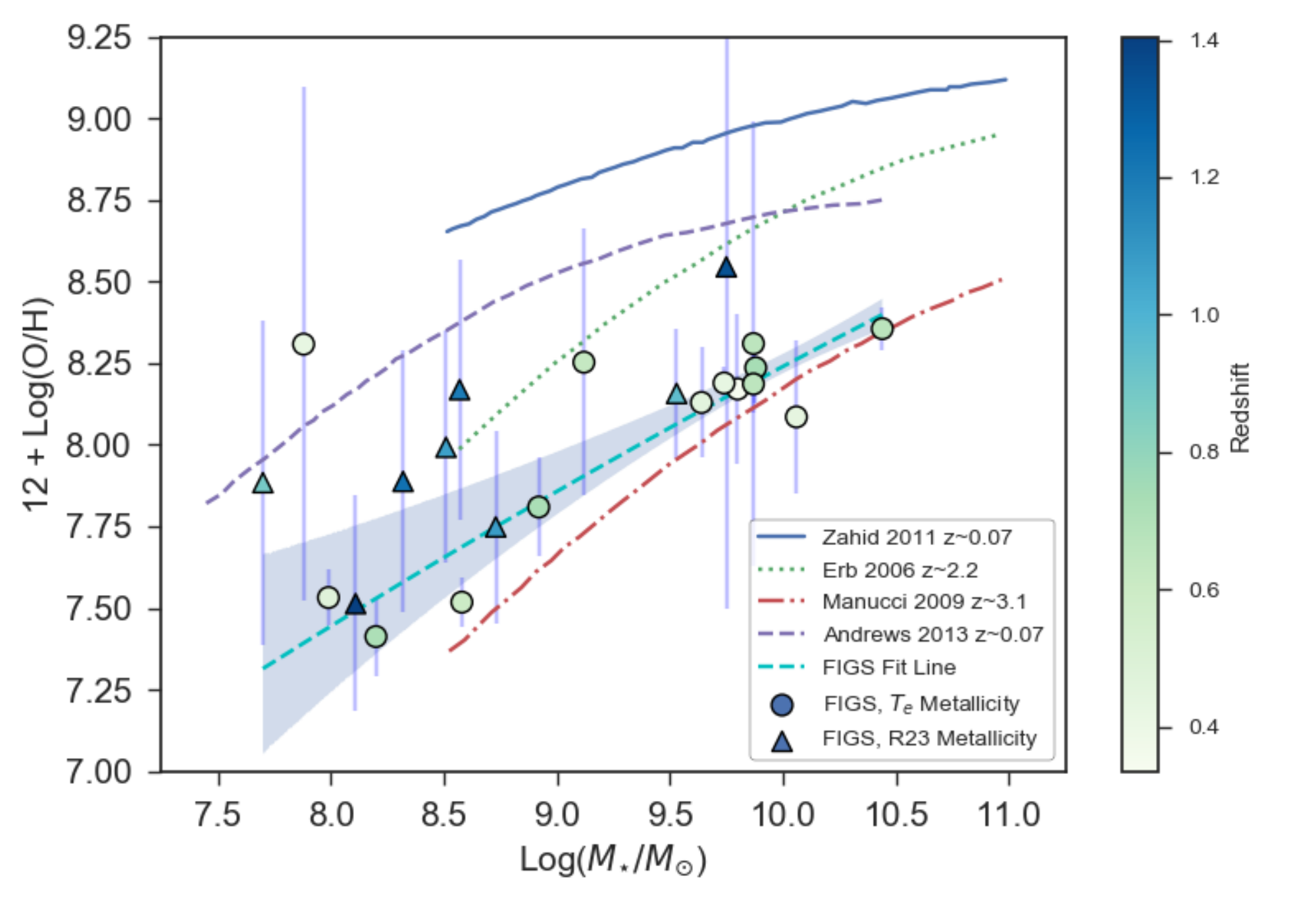}}
  \caption{The relation between the stellar masses as measured by \citet{sant15} and the gas-phase oxygen abundances as measured by $[O\textsc{ iii}]\lambda 4363$. The FIGS objects are given by circles (direct metallicity measurement) and triangles (calibrated $R_{23}$) with blue errorbars in metallicity. The FIGS objects are shaded by redshift according to the colorbar to the right. The blue solid line represents the $z \sim 0.07$ relation for local galaxies described in \citet{zah11}; the purple dashed line represents the $z \sim 0.07$ direct-metallicity relation from \citet{am13}; the green dotted line \citet{erb06} and the red dash-dot line \citet{man09} represent $R_{23}$-derived relations at $z=2.2$ and $z=3.1$.  The blue dashed line is the non-linear least squares fit to the FIGS objects, using SciPy's curve-fit function (See Equation 7 for parameters) \citep{jon01}. The blue shaded region is the $1 \sigma$ uncertainty of the fit as measured from a Monte Carlo of the $1 \sigma$ uncertainties in the fit parameters.}
\end{figure*}

\subsection{Comparison with Other MZ Relations}

Figure 6 shows the FIGS-MUSE mass-metallicity relation for the $T_e$-measured objects plotted alongside mass-metallicity relations from other surveys at similar redshift. Our measurements are offset to lower metallicity by $\sim0.6-0.7$ dex compared to these surveys. The curve from \citet{zah11} is fit from stacks of DEEP2 objects at $z\sim0.8$, for which metallicities were derived using the $R_{23}$ method. \citet{zah11} notes that since $[N\textsc{ ii}]/H\alpha$ measurements were not available, they were unable to break the $R_{23}$ degeneracy and instead assume the metallicities lie on the $R_{23}$ upper branch, though they observe that this assumption breaks down at $M_{\star} < 10^9 M_{\odot}$. \citet{sav05} derived an MZ relation for 56 $0.4<z<1.0$ galaxies from the Gemini Deep Deep Survey and the Canada-France Redshift Survey, also using the $R_{23}$ upper branch for metallicity. As described in \S4, when we apply the $R_{23}$ calculation to the 14 FIGS objects, both methods place all of them on the low-Z branch, which is itself enough to alter the metallicity measurement by up to $\sim$1 dex, enough to explain the offset in metallicity between the two surveys.

A more ``direct'' MZ comparison can be made from the \citet{am13} (hereafter AM13) result, shown in Figure 7 as a purple dashed curve. The AM13 MZ relation is derived from stacks of direct-method metallicity calculations of local SDSS galaxies at $z=0.07$. The direct-metallicity FIGS measurements are denoted by circles and the calibrated-$R_{23}$ measurements denoted by triangles. Despite also using the [O\textsc{ iii}]4363 $T_{e}$ method, AM13 find higher metallicities than we find, with a median metallicity offset of +0.65 dex. 

\subsection{Discussion of the Offset}

In this section, we examine possible causes for the low-metallicity offset of our sample.

\subsubsection{Redshift}

In Figures 6 and 7, the FIGS points are colored according to the line-centroid-derived redshift, but no significant trend in redshift emerges from among these 14 ELGs. This agrees with the results of \citet{sav05}, who also found no significant redshift evolution in metallicity in their sample at a similar redshift range.

 The median redshift in the AM13 sample is $z=0.07$, with a maximum of $z=0.25$. This is lower than the minimum redshift in the FIGS sample ($z=0.371$), and the median redshift in the FIGS $T_e$-derived sample is $z=0.614$. Previous surveys \citep{mai08, zah13} suggest the metallicity evolution from $z=0$ to $z\sim 0.8$ is roughly 0.1-0.2 dex at a given stellar mass. This is not large enough to account for the offset between FIGS and AM13, though possibly the $R_{23}$ measurements used by the previous surveys underestimate this evolution. This offset does allow for the FIGS objects to fall within the scatter of the metal-poor galaxies in the AM13 sample. 

\citet{jon15} selected a sample of 32 DEEP2 galaxies with [O\textsc{ iii}]4363 emission at $ z\sim 0.8$ from which they calculated gas-phase metallicities in the range $7.8 < 12 + \log (O/H) < 8.4$. They do not include a mass-metallicity relation, but most of the FIGS objects have metallicities that are consistent with this metallicity range to within the $1\sigma$ error. Of the three objects with significantly lower measured metallicity, only one is at a redshift at the higher end of the sample redshift ($z=0.719$), and thus at a comparable redshift to the Jones sample, and all three are at relatively low mass. 

\subsubsection{IMF}

Inconsistency in the IMF used to derive stellar masses for different studies can produce offsets in stellar mass, which in turn would affect the MZ Relation. Masses for the \citet{zah11} relation were also caulculated using a Chabrier IMF, while the \citet{sav05} relation uses masses with an IMF derived by \citet{bg03} that produces masses 1.2 times smaller than Kroupa. A calculation of IMF offsets \citep{zah12} suggests an offset of +0.03 dex between Kroupa-derived and Chabrier-derived stellar masses, and an offset of -0.07 dex between Chabrier masses and those used in \citet{sav05}. The \citet{am13} relation used masses derived from a Kroupa IMF, which should result in a +0.03 mass offset compared to the Chabrier masses used in the FIGS relation. These offsets are all comparable to the $<0.1$ dex scatter in stellar masses in the \citet{san15} catalogs, and are much too small to explain the metallicity offset.

\subsubsection{Contributions of Multiple HII Regions}

Another explanation for the low metallicities we measure is the possibility that the lines we detect are dominated by emission from particularly extreme regions within the galaxy. An HII region with an especially low metallicity and large electron temperature could produce stronger [OIII]4363 emission for that region. In a small galaxy, the flux from such a region could dominate compared to flux from milder regions, resulting in that region's low metallicity measurement reducing the overall metallicity measurement for the galaxy \citep{san17}. This could perhaps explain the extremely low metallicities of the lowest-mass objects, but does not account for the lower M-Z relation overall.

\subsubsection{Selection Effects, Line Emission, and Star Formation}

In Figure 7, we have also included the FIGS objects with calibrated $R_{23}$ metallicities (see \S4), denoted by triangles. There continues to be no significant redshift evolution, as these new, higher-z objects tend to have higher metallicity. This is likely a selection effect: the highest redshift objects are also exclusively $R_{23}$-calibrated. This means that there cannot have been a detected [O\textsc{ iii}]4363 line, which itself implies possibly lower [O\textsc{ iii}]4363 emission, which in turn implies a higher metallicity for the objects in the calibration sample. Furthermore, these objects are typically fainter, resulting in larger flux errors which contribute to broader error bounds on the metallicities. The high-z objects are still consistent with the possible range of the MZ fit, and there are a few objects in the $R_{23}$ sample where we measure low metallicities comparable to what we measure with [O\textsc{ iii}]4363. This means that [O\textsc{ iii}]4363 selection alone cannot fully account for the metallicity offsets, and so lends support to the findings in \citet{xia12}, which suggest that emission line strength itself is an indicator of low metallicity.

More recently, \citet{amo17} find a sample of $2.4 < z < 3.5$ galaxies with $7.4 < 12 + \log(O/H) < 7.7$ using strong UV emission lines. They find low metallicities consistently across a broad range of stellar masses, up to $\log(M_{\star}/M_{\odot}) = 9.8$. Their sample also shows indicators of recent star formation, suggesting a link between star formation and metallicity somewhat independent of the stellar mass. We explore the effects of star formation for the FIGS objects in the following section.

\subsection{SFR and the Fundamental Metallicity Relation}

We calculate the star formation rate (SFR) for the 14 objects with direct-measurement metallicities based on the line flux conversion given in \citet{ken98}

\begin{equation}
  SFR(M_{\odot} \text{year}^{-1}) = 7.9 \cdot 10^{-42} L(H\alpha) (\text{erg s}^{-1})
\end{equation}

using the extinction-corrected $H\alpha$ line flux. The SFR error is estimated based on the line flux error. The metallicity as a function of SFR is shown in Figure 8, along with a non-linear fit. This shows a trend of metallicity increasing with the SFR. In Figure 9, we plot the gas-phase metallicity versus the Specific Star Formation Rate (SSFR), which is the SFR per stellar mass, as we as the fit of the FIGS objects. This shows a slight trend of declining metallicity with increased SSFR, with the lowest metallicity (and smallest mass) galaxies having SSFR $> 10^{-9}$ yr$^{-1}$. \citet{ell08} has shown a relation where metallicity is lower for galaxies with higher SSFR at a given stellar mass, with a metallicity offset of up to 0.15 dex at the lowest stellar masses ($M_{\star} \sim 10^{8.5} M_{\odot}$ in their study). This suggests that the large SSFR we observe in several of the FIGS objects could be a driver for the low-metallicity offset compared to other mass-metallicity relations. If this is the case, it likely has implications for how star formation interacts with the nebular gas. There are two plausible scenarios. First, inflows of circumgalactic gas could both bring lower-metallicity gas into the galaxy and trigger new star formation, producing strong line emission in the metal-deficient medium around the new stars. Alternatively, recent star formation in a galaxy produces strong stellar winds and supernovae, which could cause outflows that push the most metal-enriched gas out of the galaxy. In either case, increased star formation would show a clear link with measuring reduced metallicity in a galaxy's nebular gas.

\begin{figure}
  \includegraphics[width=0.5\textwidth]{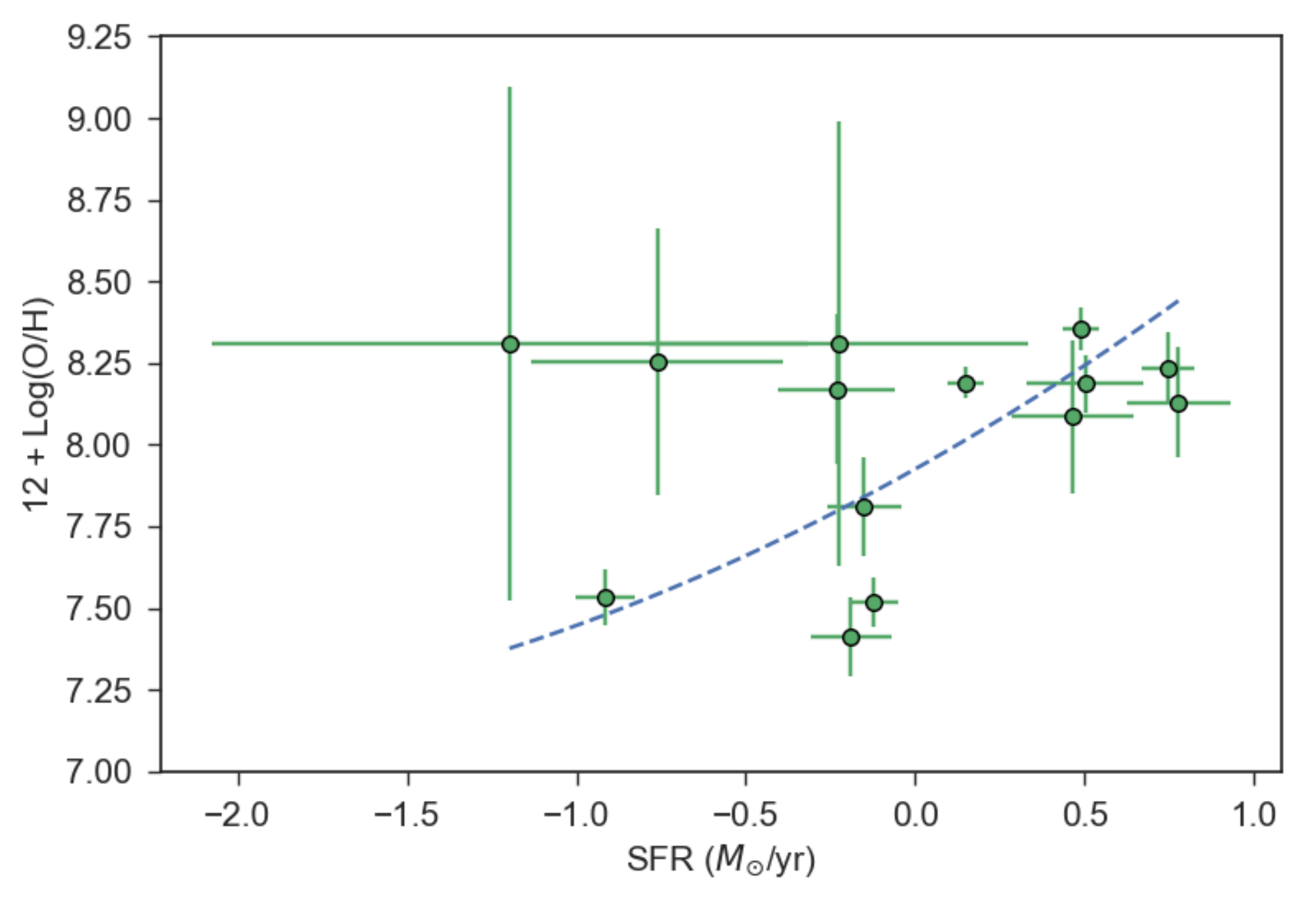}
  \caption{The gas-phase metallicity of the FIGS objects as a function of the SFR. The dashed line shows the non-linear least-squares fit, which shows a trend of increasing metallicity with increasing SFR.}
\end{figure}

\begin{figure*}
  \includegraphics[width=\textwidth]{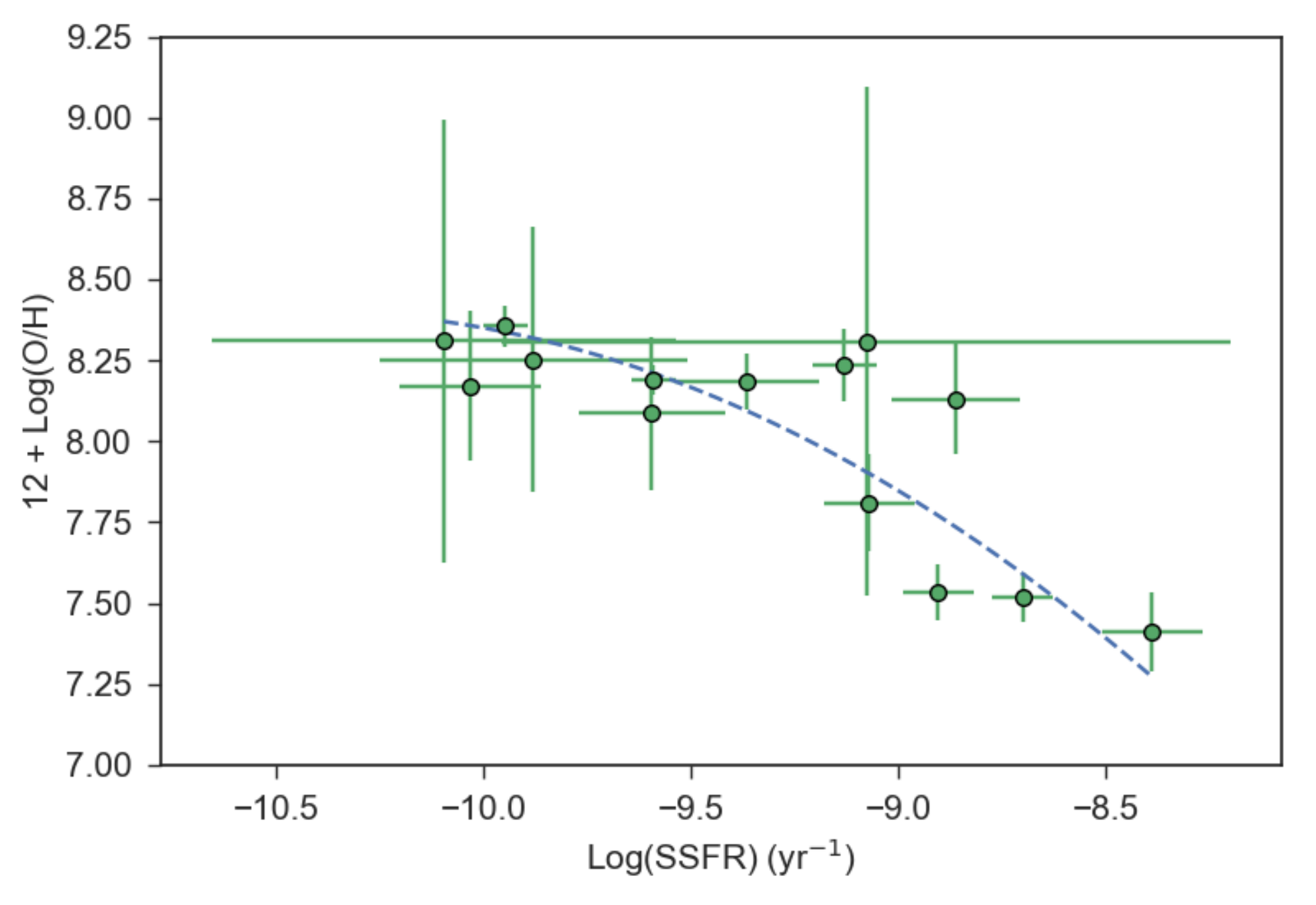}
  \caption{The gas-phase metallicity as a function of the specific star formation rate (SSFR). The dashed line shows the non-linear least-squares fit of the FIGS points, which shows a possible trend of lower metallicity at increased SSFR.}
\end{figure*}

Further investigation of the relationship between these parameters is needed. \citet{man10} describes the dependence of the gas-phase metallicity on stellar mass and the SFR as the Fundamental Metallicity Relation, for which they obtain the projection

\begin{equation}
  12 + \log \left( \frac{O}{H} \right) = \log \left( \frac{M_{\star}}{M_{\odot}} \right) - 0.32 \cdot \log (SFR).
\end{equation}

This projection, derived from a sample of SDSS $z=0.07-0.30$ ELGs which had an H$\alpha$ $S/N > 25$, minimizes the scatter in metallicity around the relation. Mannucci et al. also find good agreement with the FMR and this projection for galaxies up to $z=2.5$. We calculated this projected FMR for the FIGS galaxies with SFR, plotted in Figure 10, to see how well our results match this relation. The FIGS 14 objects follow the trend of the lower FMR, but sit lower on the plot due to their lower metallicities. This is perhaps partially accounted for by the differences between direct and $R_{23}$ metallicity measurements as described in \S5.1, though the metallicities used in Mannucci et al. were derived from either $R_{23}$ or from the [N\textsc{ ii}]$\lambda$6584/H$\alpha$ ratio. Mannucci et al. estimates a 0.09 dex offset in metallicity between these two methods, making the magnitude of the offset from $R_{23}$ and direct measurements difficult to determine. Figure 10 also shows a difference in the range of values for the $M_{\star}$-SFR axis, with the FIGS sample probing much lower stellar masses than (and thus also lower SSFR than) the Mannucci sample. While the FIGS objects span a range of SFR similar to that seen in Mannucci et al., the $M_{\star}$ values are lower, and we do not know how well Mannucci's projection reduces scatter at lower stellar mass.

We also tried comparing our results to the Fundamental Plane of Metallicity (FPZ) derived by \citet{hun16} using the Metallicity Evolution and Galaxy Assembly (MEGA) data set. Hunt et al. attempted to derive a fundamental relation between metallicity, mass, and SFR from a large set of galaxies with a wide range of properties and redshifts, including a variety of strong-line methods for measuring the metallicity (the direct method among them, but not predominantly so). With this data set, Hunt et al. performed a Principal Component Analysis to derive a plane relating the three variables:

\begin{equation}
12 + \log \left( \frac{O}{H} \right) = -0.14 \log(SFR) + 0.37 \log \left( \frac{M_{\star}}{M_{\odot}} \right) + 4.82
\end{equation}

In Figure 11, we plotted the FIGS objects on this plane. The blue line gives the one-to-one correspondence given by Equation 10, with the shaded region providing the $\sigma=0.16$ scatter from Hunt et al.'s narrowest residual distribution. The FIGS points lie systematically below this, though within the total scatter of MEGA objects around it. The dashed and dot-dashed lines show linear fits to the FIGS points: the red (dashed) line allows both parameters of the linear fit to move freely, while the purple (dot-dash) line assumes the same slope as the one-to-one correspondence and only lets the y-intercept vary. This produces an overall metallicity offset of $\sim 0.3$ dex. The linear fit demonstrates that the higher-metallicity FIGS ELGs are actually fairly consistent with $1\sigma$ range of the FPZ measure, and that the lowest-metallicity objects are the ones driving the offset.

This all suggests that our sample of galaxies with direct metallicity measurements includes some uniquely low-mass, low-metallicity objects.

\begin{figure*}
  \includegraphics[width=\textwidth]{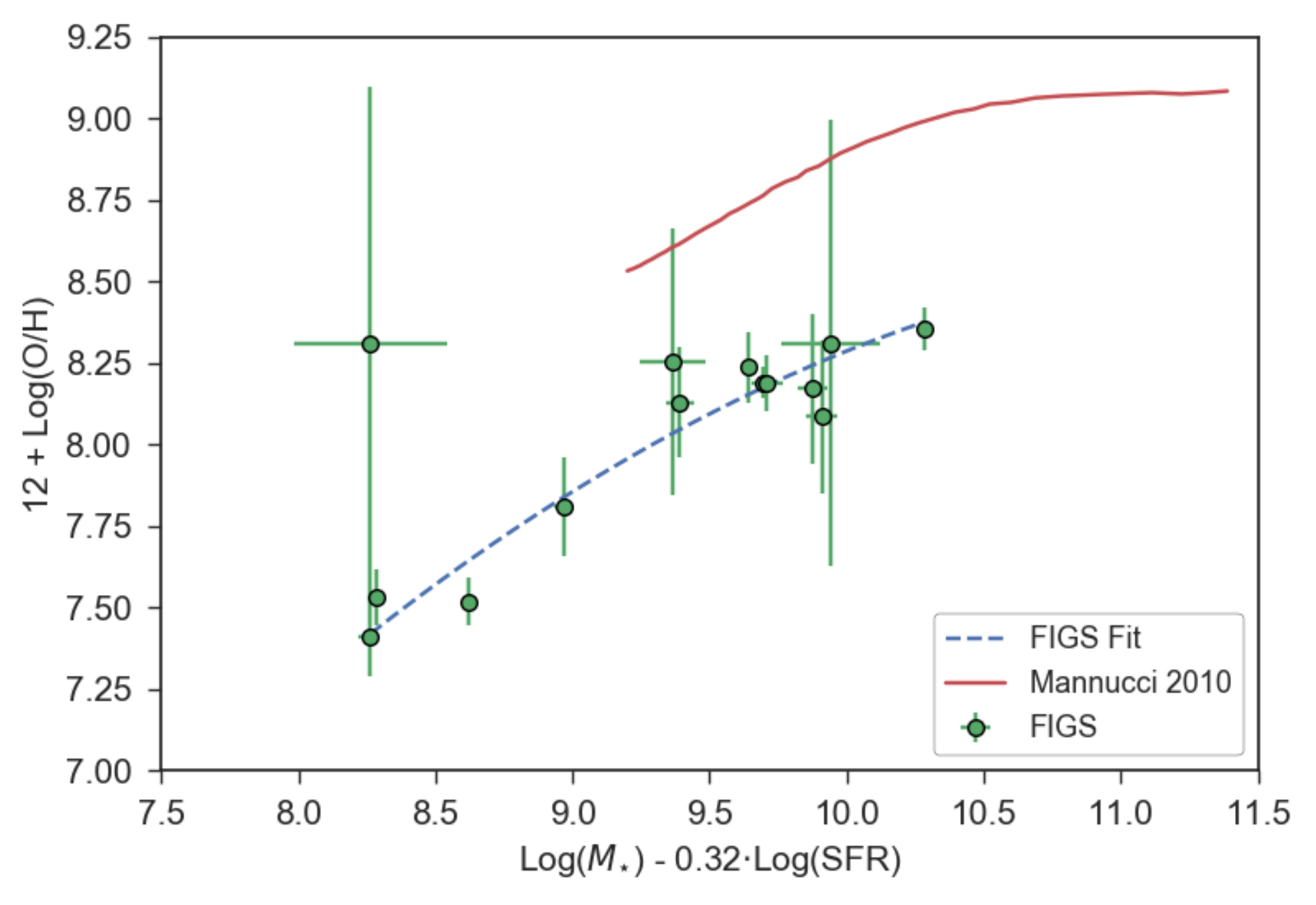}
  \caption{The gas-phase metallicity as a function of the Fundamental Metallicity Relation (FMR), as given by \citet{man10}. The Mannucci relation is shown by the red solid line, with the FIGS points in green. The blue dashed line gives the non-linear least-squares fit of the FIGS points.}
\end{figure*}

\begin{figure*}
\includegraphics[width=\textwidth]{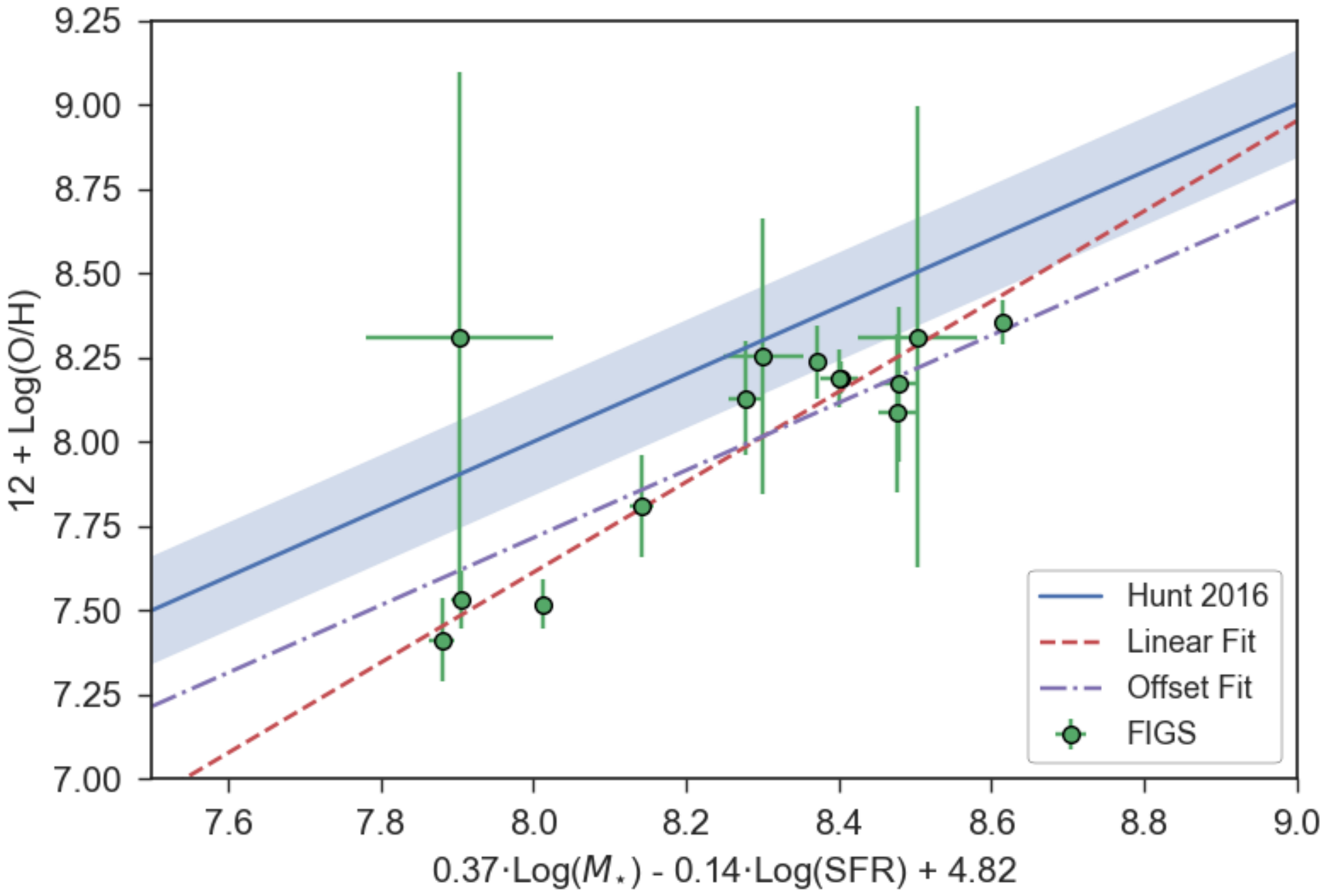}
\caption{This plot shows the 14 direct-method FIGS objects plotted using the Fundamental Plane of Metallicity formulation from \citet{hun16}. The blue line shows the one-to-one correlation of metallicity to combined mass and SFR around which the formulation was based. The FIGS points are in green, and the dashed red line represents the simple linear fit to the FIGS data. The dot-dashed purple line shows the linear fit to the FIGS data if the slope is fixed to match the Hunt correlation. This results in a metallicity offset of 0.28.}
\end{figure*}

\section{Conclusions}

By using near-infrared spectroscopy from FIGS, we were able to identify 71 ELGs in the GS1/HUDF field, primarily through the identification of H$\alpha$, [O\textsc{ iii}]$\lambda$5007, and [O\textsc{ ii}]$\lambda$3727 emitters in the redshift range of $0.3 < z < 2.0$. We were able to confirm 41 out of the 71 ($\sim 58\%$) by identifying complementary lines in matching optical data, either with ACS grism spectroscopy from the previous GRAPES survey, or from new MUSE-VLT optical spectroscopy. We measure line fluxes down to a sensitivity of $10^{-17}$ ergs cm$^{-2}$ s$^{-1}$ in FIGS and $\sim 3 \cdot 10^{-19}$ ergs cm$^{-2}$ s$^{-1}$ from MUSE-VLT.

Out of these objects, we found 14 for which we were able to measure the auroral [O\textsc{ iii}]4363 emission line in MUSE optical spectra with a S/N ratio of at least 3, with a redshift range of $0.3 < z < 0.8$ and with stellar masses down to $10^{7.9} M_{\odot}$. We used these measurements to calculate the gas-phase metallicity via the electron temperature, and from this we produced a mass-metallicity relation. When compared to MZ relations at similar redshifts, we find a significant offset to lower metallicity. We examined several possible causes for the offset, and find that redshift evolution does not account for the difference in metallicity. The offset can be only partially explained by differences with this metallicity derivation method compared to the more common $R_{23}$ method, as previously seen in \citet{am13, san16}. Selecting [O\textsc{iii}]4363 line emitters does select for lower metallicity in general, but with a new $R_{23}$ calibration we found other galaxies at similarly low metallicity, so selection effects alone cannot account for the difference. To further explore the metallicity offset, we determined the SFR, SSFR, and FMR for the sample. We find a trend between metallicity and SSFR, showing that the low-mass, low-metallicity FIGS objects have a large SSFR and are low-metallicity outliers in attempts to find a fundamental relationship between these parameters. This suggests that recent star formation is connected to inflows or outflows of nebular gas, leading to the measurement of low-metallicity gas in the galaxy. The existence of such outliers shows the need for further spectroscopic analysis of low-mass galaxies, which may be host to significant activity well after the universal peak of star formation at $z \simeq 2$.

{}

\begin{table*}
\begin{center}
\caption{All observed emission line fluxes for 71 galaxies in GS1/HUDF}
\begin{tabular}{ccccccccccc}
\tableline
FIGS ID & RA & Dec & F105W &  H$\alpha$ & [OIII] & H$\beta$ & [OIII]4363 & [OII]3727 & z\footnotemark[3] \\
\tableline
724 & 53.17226 & -27.76062 & 22.56 & - & - & - & - & $532.6\pm53.3$ & 1.550 \\
950 & 53.16150 & -27.76762 & 23.43 & $788.8\pm78.9$ & $254.4\pm31.3$\footnotemark[2] & $101.3\pm14.4$\footnotemark[2] & $17.23\pm8.76$\footnotemark[2] & $259.6\pm29.8$\footnotemark[2] & 0.678 \\ 
970 & 53.16018 & -27.76931 & 24.06 & - & $621.2\pm36.0$ & $138.8\pm16.9$ & - & $216.4\pm51.9$\footnotemark[2] & 1.037 \\ 
1013 & 53.16993 & -27.77103 & 19.97 & $772.1\pm77.2$ & $368.0\pm42.2$\footnotemark[2] & $674.3\pm95.8$\footnotemark[2] & - & $1362\pm109$\footnotemark[2] & 0.622 \\
1016 & 53.17210 & -27.77038 & 23.62 & $215.3\pm21.5$ & $49.7\pm12.0$\footnotemark[2] & $33.6\pm23.4$\footnotemark[2] & $17.1\pm9.2$\footnotemark[2] & $88.5\pm12.5$\footnotemark[2] & 0.622 \\  
1056 & 53.16245 & -27.77091 & 24.30 & - & $260.7\pm13.4$ & $33.2\pm3.3$ & - & $43.1\pm10.3$\footnotemark[2] & 1.038 \\
1103 & 53.17400 & -27.77206 & 20.72 & $2019\pm285$\footnotemark[1] & $80.0\pm17.7$\footnotemark[2] & $178.2\pm20.3$\footnotemark[2] & $70.64\pm17.81$\footnotemark[2] & $302.6\pm30.1$\footnotemark[2] & 0.334 \\ 
1132 & 53.18448 & -27.77225 & 24.58 & - & $197.5\pm42.5$\footnotemark[1] & $59.8\pm36.4$\footnotemark[1] & - & $47.8\pm24.4$\footnotemark[2] & 0.840 \\ 
1171 & 53.15122 & -27.77284 & 23.79 & $866.7\pm86.7$ & $756.8\pm43.5$\footnotemark[2] & $158.4\pm22.0$\footnotemark[2] & $21.1\pm11.3$\footnotemark[2] & $331.8\pm30.0$\footnotemark[2] & 0.606 \\
1239 & 53.19146 & -27.77389 & 23.48 & $666.9\pm66.7$ & - & - & - & - & 0.420 \\
1295 & 53.16236 & -27.77506 & 20.58 & $1869\pm522$\footnotemark[1] & $198.4\pm21.4$\footnotemark[2] & $476.2\pm49.1$\footnotemark[2] & $108.9\pm46.0$\footnotemark[2] & $655.0\pm54.8$\footnotemark[2] & 0.420 \\
1296 & 53.15936 & -27.77503 & 23.09 & - & $354.3\pm31.9$ & $131.6\pm13.2$ & - & - & 1.219 \\
1299 & 53.16080 & -27.77537 & 21.24 & $657.2\pm65.7$ & $126.0\pm76.3$\footnotemark[2] & - & $18.1\pm8.9$\footnotemark[2] & $671.0\pm48.4$\footnotemark[2] & 0.622 \\
1316 & 53.16531 & -27.77486 & 27.01 & - & $91.6\pm8.0$ & $77.2\pm7.7$ & - & $79.4\pm31.6$\footnotemark[2] & 1.253 \\ 
1359 & 53.18591 & -27.77561 & 22.90 & - & - & - & - & $240.3\pm76.7$\footnotemark[1] & 1.414 \\
1392 & 53.18105 & -27.77618 & 22.08 & $1335\pm71.2$ & $150.6\pm52.4$\footnotemark[2] & $72.0\pm9.2$\footnotemark[2] & $31.4\pm10.3$\footnotemark[2] & $211.0\pm29.1$\footnotemark[2] & 0.668 \\ 
1467 & 53.15105 & -27.77731 & 24.14 & $355.3\pm21.1$ & $336.6\pm50.6$\footnotemark[1] & $69.0\pm38.6$\footnotemark[2] & - & $194.1\pm25.9$\footnotemark[2] & 0.736 \\
1476 & 53.14744 & -27.77760 & 23.63 & - & - & - & - & $189.9\pm11.3$ & 1.859 \\
1477 & 53.15829 & -27.77745 & 24.49 & - & - & - & - & $310.6\pm19.6$ & 1.556 \\
1481 & 53.14661 & -27.77749 & 25.03 & - & $281.3\pm21.9$ & $56.7\pm6.0$ & - & $106.7\pm27.5$\footnotemark[2] & 1.088 \\
1500 & 53.15234 & -27.77795 & 24.32 & - & - & - & - & $214.0\pm74.1$\footnotemark[1] & 1.413 \\
1552 & 53.15720 & -27.77852 & 23.83 & - & - & $78.0\pm12.7$ & - & $299.0\pm44.3$\footnotemark[1] & 1.307 \\
1689 & 53.16248 & -27.78035 & 25.11 & $559.4\pm48.9$ & $707.6\pm83.1$\footnotemark[1] & $103.0\pm32.4$\footnotemark[2] & $31.23\pm14.41$\footnotemark[2] & $39.7\pm10.5$\footnotemark[2] & 0.719 \\  
1711 & 53.19700 & -27.78060 & 23.86 & - & $575.4\pm45.5$ & - & - & - & 0.733 \\ 
1728 & 53.17633 & -27.78086 & 24.99 & $684.8\pm44.5$ & $660.2\pm43.9$\footnotemark[1] & $89.6\pm10.9$\footnotemark[2] & - & $109.1\pm15.7$\footnotemark[2] & 0.535 \\
1803 & 53.17007 & -27.78207 & 26.68 & - & - & $56.2\pm7.5$ & - & $134.7\pm27.2$\footnotemark[1] & 1.351 \\
1829 & 53.15076 & -27.78256 & 24.22 & - & - & $90.4\pm9.0$ & - & $136.6\pm38.1$\footnotemark[1] & 1.352 \\
1851 & 53.15278 & -27.78270 & 24.44 & - & $926.3\pm100.5$\footnotemark[1] & $124.7\pm20.9$\footnotemark[2] & - & $260.4\pm23.5$\footnotemark[2] & 0.764 \\
1900 & 53.18457 & -27.78332 & 24.42 & - & $979.0\pm69.8$ & $100.5\pm10.1$ & - & $195.6\pm27.7$\footnotemark[2] & 1.136 \\ 
1901 & 53.18433 & -27.78337 & 25.78 & - & $254.9\pm18.6$ & $13.8\pm1.4$ & - & - & 1.257 \\  
1946 & 53.19259 & -27.78379 & 24.71 & - & $429.1\pm32.5$ & $42.1\pm7.6$ & - & - & 0.869 \\
2023 & 53.15186 & -27.78475 & 25.58 & - & $285.4\pm25.9$ & $36.2\pm3.6$ & - & $36.2\pm18.2$\footnotemark[2] & 1.219 \\
2039 & 53.16657 & -27.78486 & 27.62 & - & - & $29.6\pm5.2$ & - & $102.7\pm27.0$\footnotemark[1] & 1.320 \\
2049 & 53.16935 & -27.78499 & 27.13 & - & - & $27.3\pm2.7$ & - & $100.3\pm26.5$\footnotemark[1] & 1.344 \\
2138 & 53.16048 & -27.78630 & 24.31 & - & $356.4\pm28.9$ & $66.5\pm6.5$ & - & - & 0.984 \\
2168 & 53.16347 & -27.78664 & 25.14 & $284.9\pm19.6$ & $227.7\pm18.9$\footnotemark[2] & $41.5\pm3.9$\footnotemark[2] & $13.73\pm4.08$\footnotemark[2] & $64.2\pm5.7$\footnotemark[2] & 0.468 \\ 
2187 & 53.17775 & -27.78697 & 24.37 & - & $336.3\pm21.7$ & $99.7\pm11.0$ & - & $143.2\pm22.8$\footnotemark[2] & 0.955 \\
2221 & 53.16410 & -27.78730 & 23.78 & - & $2168\pm156$ & $318.2\pm31.8$ & - & $284.1\pm50.1$\footnotemark[2] & 1.097 \\ 
2291 & 53.14930 & -27.78853 & 23.05 & - & - & - & - & $453.0\pm45.3$ & 1.917 \\
2338 & 53.15736 & -27.78922 & 25.11 & - & $346.1\pm11.6$ & $3.6\pm0.4$ & - & $18.7\pm17.8$\footnotemark[2] & 1.015 \\
2363 & 53.16802 & -27.78967 & 22.80 & $717.5\pm61.3$ & $373.2\pm30.6$\footnotemark[2] & $30.9\pm20.9$\footnotemark[2] & - & $626.8\pm48.2$\footnotemark[2] & 0.619 \\
2378 & 53.18795 & -27.79000 & 20.25 & $8916\pm568$ & $198.1\pm36.0$\footnotemark[2] & $489.6\pm44.1$\footnotemark[2] & $81.29\pm17.89$\footnotemark[2] & $640.2\pm56.5$\footnotemark[2] & 0.436 \\
2385 & 53.18481 & -27.78993 & 23.13 & - & $392.3\pm31.2$ & $160.0\pm25.4$ & - & $431.7\pm62.6$\footnotemark[2] & 0.954 \\
2417 & 53.16042 & -27.79037 & 23.48 & - & - & - & - & $354.6\pm35.5$ & 1.614 \\
2495 & 53.18413 & -27.79153 & 23.07 & - & $251.2\pm14.5$ & $50.6\pm5.1$ & - & - & 1.224 \\
2517 & 53.16161 & -27.79230 & 20.62 & $16439\pm1374$ & $3238\pm198$\footnotemark[2] & $1127\pm88$\footnotemark[2] & $45.2\pm17.46$\footnotemark[2] & $3279.0\pm202.2$\footnotemark[2] & 0.459 \\ 
2560 & 53.18416 & -27.79264 & 21.41 & $5114\pm225$ & $1810\pm226$\footnotemark[1] & $716\pm54$\footnotemark[2] & $146.72\pm54.11$\footnotemark[2] & $1495\pm101$\footnotemark[2] & 0.738 \\
2570 & 53.16412 & -27.79265 & 27.02 & - & - & $30.9\pm3.1$ & - & $89.6\pm25.7$\footnotemark[1] & 1.311 \\
2654 & 53.18221 & -27.79399 & 24.81 & $278.9\pm27.9$ & $996.6\pm53.5$\footnotemark[1] & $33.3\pm28.8$\footnotemark[2] & - & $80.6\pm16.4$\footnotemark[2] & 0.734 \\
2669 & 53.15663 & -27.79430 & 24.33 & - & $329.3\pm32.3$ & $118.8\pm11.9$ & - & - & 1.094 \\
2696 & 53.15586 & -27.79490 & 22.94 & - & $785.2\pm42.9$ & $215.5\pm17.6$ & - & - & 1.104 \\
2720 & 53.15675 & -27.79558 & 21.80 & - & $731.9\pm35.8$ & $330.3\pm33.0$ & - & $398.5\pm76.1$\footnotemark[2] & 1.099 \\ 
2732 & 53.16133 & -27.79580 & 23.59 & - & - & - & - & $485.7\pm134.9$\footnotemark[1] & 1.498  \\ 
2783 & 53.18808 & -27.79574 & 24.03 & $125.6\pm12.6$ & $43.3\pm16.6$\footnotemark[2] & - & $9.3\pm6.8$\footnotemark[2] & $15.4\pm10.0$\footnotemark[2] & 0.532 \\ 
2872 & 53.16687 & -27.79771 & 23.68 & - & $615.7\pm42.9$ & $66.3\pm6.6$ & - & $409.5\pm28.0$\footnotemark[2] & 0.984 \\
2942 & 53.16112 & -27.79880 & 25.53 & - & $387.9\pm34.1$ & $19.5\pm1.9$ & - & $231.4\pm90.6$\footnotemark[2] & 1.238 \\
4198 & 53.17838 & -27.76824 & 20.22 & $1964.6\pm115.3$ & $317.8\pm276.7$\footnotemark[2] & $474.1\pm184.8$\footnotemark[2] & $110.9\pm16.5$\footnotemark[2] & $1073.9\pm97.2$\footnotemark[2] & 0.669 \\ 
4258 & 53.15229 & -27.77009 & 23.66 & - & - & - & - & $420.6\pm32.9$ & 1.859 \\ 
4284 & 53.18454 & -27.76822 & 25.22 & - & - & - & - & $74.4\pm7.4$ & 1.839 \\
6865 & 53.19033 & -27.77430 & 26.84 & - & $195.7\pm12.1$ & $119.7\pm12.0$ & - & - & 0.902 \\
8178 & 53.18766 & -27.78378 & 26.95 & $84.3\pm8.4$ & $126.8\pm37.2$\footnotemark[1] & $91.9\pm44.5$\footnotemark[2] & - & $16.8\pm15.3$\footnotemark[2] & 0.739
\end{tabular}
\scriptsize {
  \begin{tablenotes}
  Fluxes are given in units of $10^{-19}$ ergs s$^{-1}$ cm$^{-2}$ \\
  \item[]$\rm ^a$ Line measured in FIGS and MUSE. \\
  \item[]$\rm ^b$ Line measured only in MUSE. \\
  \item[]$\rm ^c$ Derived from central wavelength of the most significantly detected line, averaged from PAs and MUSE. 
  \end{tablenotes}}
\end{center}
\end{table*}

\end{document}